\documentclass[useAMS,usenatbib, times]{mn2e}
\usepackage{latexsym,graphicx,natbib,placeins}
\usepackage{placeins}
\usepackage{amsmath,amssymb}
\usepackage[toc,page]{appendix}
\newcommand{\unit}[1]{\, \rm{#1}}	
\newcommand{\sscript}[1]{{_{\rm{#1}}}}		

\title[A semi-analytic dynamical friction model that reproduces core-stalling]{A semi-analytic dynamical friction model that reproduces core-stalling}
\author[J. A. Petts, A. Gualandris, J. I. Read]{J. A. Petts$^{1}$\thanks{E-mail:
j.petts@surrey.ac.uk}, A. Gualandris, J. I. Read\\
$^{1}$University of Surrey, Department of Physics, Guildford, United Kingdom}
\begin{document}

\date{Accepted 24 September 2015}

\pagerange{\pageref{firstpage}--\pageref{lastpage}} \pubyear{2015}

\maketitle

\label{firstpage}

\begin{abstract}
We present a new semi-analytic model for dynamical friction based on Chandrasekhar's formalism. The key novelty is the introduction of physically motivated, radially varying, maximum and minimum impact parameters. With these, our model gives an excellent match to full N-body simulations for isotropic background density distributions, both cuspy and shallow, without any fine-tuning of the model parameters. In particular, we are able to reproduce the dramatic core-stalling effect that occurs in shallow/constant density cores, for the first time. This gives us new physical insight into the core-stalling phenomenon. We show that core stalling occurs in the limit in which the product of the Coulomb logarithm and the local fraction of stars with velocity lower than the infalling body tends to zero. For cuspy backgrounds, this occurs when the infalling mass approaches the enclosed background mass. For cored backgrounds, it occurs at larger distances from the centre, due to a combination of a rapidly increasing minimum impact parameter and a lack of slow moving stars in the core. This demonstrates that the physics of core-stalling is likely the same for both massive infalling objects and low-mass objects moving in shallow density backgrounds. We implement our prescription for dynamical friction in the direct summation code NBODY6 as an analytic correction for stars that remain within the Roche volume of the infalling object. This approach is computationally efficient, since only stars in the inspiralling system need to be evolved with direct summation. Our method can be applied to study a variety of astrophysical systems, including young star clusters orbiting near the Galactic Centre; globular clusters moving within the Galaxy; and dwarf galaxies orbiting within dark matter halos.
\end{abstract}

\begin{keywords}
Galaxy: kinematics and dynamics -- Galaxies: star clusters -- methods: numerical.
\end{keywords}

\section{Introduction}

In a seminal work, \citet{b1} showed that massive objects moving through an infinite homogeneous stellar medium will experience a drag force that he called `dynamical friction' \citep[see][and \S\ref{theory.ch}]{b2}. This frictional force is likely responsible for galactic mergers \citep{b31,b32}, the accretion of satellites and globular clusters onto the Galaxy \citep{b31,b17}, and even (in part) the coalescence of supermassive black holes \citep{b40}.

Numerical simulations using both direct summation codes and collisionless codes have shown that Chandrasekhar's formula (equation \ref{dynfric.eq}) works remarkably well, despite the fact that it likely misses important physics like resonant interactions between the infalling body and the background \citep{b27,b39,b28}. Nonetheless, there are some situations in which it has been reported to perform poorly. Most notably, in the case of a constant density background \citep{b24,b39,b35}, or when the mass of the infalling body approaches the enclosed background mass \citep{b26}. The former is perhaps surprising given that the original derivation assumes a homogeneous sea of background stars. But this failure could owe to the extreme resonance of such harmonic potentials \citep{b24}. 

While dynamical friction can be accurately modelled using N-body simulations, it can often be prohibitively expensive. To achieve accurate results, the background "stars" (we shall call them "stars", though these could be any distribution of gravitating masses, e.g. dark matter particles) must be substantially less massive than the infalling body. Otherwise, the body will simply be stochastically buffeted around, experiencing little friction \citep{b3}. If we wish to self consistently model the internal dynamics of a globular cluster, for example, falling onto the Galaxy the numerical requirements rapidly become extreme. For example a globular cluster of $10^5 M\unit{\odot}$ ($\mathcal{O}(10^5)$ particles), would need $10^{7-12}$ background particles to accurately model the inspiral in a massive host. If a lower mass resolution is used for the background, and gravitational softening is employed to keep the relaxation time the same, the dynamical friction force is under-predicted (see \S\ref{gt_discussion.ch}). For this reason, semi-analytic models of dynamical friction have become invaluable \citep{b12,b21,b22,b25}. These significantly speed up the simulations since only the internal dynamics of the satellite needs to be integrated self-consistently, and the effects of a particular background distribution are modelled in an analytic way. Such models have been well-tested for point mass satellites in steep power-law density backgrounds, giving a good match to full N-body simulations \citep{b22}. However, in shallow or constant density backgrounds, dynamical friction stalls \citep{b24} -- an effect that has so far not been captured by semi-analytic models. This has led some authors to move away from semi-analytic modelling, towards particle-mesh codes calibrated by direct $N$-body and tree-codes \citep{b4}, mixed collisional/collisionless methodologies \citep{b14,b10,b7}, and accurate advanced tree-codes \citep{b36}. While these are exciting new approaches, they remain computationally expensive. This begs the question of whether the semi-analytic models cannot be improved. Such improvements would not only open up new classes of astrophysically interesting problems, but also shed more light on the interesting physics of dynamical friction and core stalling. This is the goal of this present work.

In this paper, we introduce a new semi-analytic model for dynamical friction based on the familiar Chandrasekhar formalism (equation \ref{dynfric.eq}). Our key novelty is that we present a new physically motivated, and radially varying minimum impact parameter ($b\sscript{min}$), which when combined with the maximum impact parameter ($b\sscript{max}$) from \citet{b21}, gives our semi-analytic model a remarkable match to a large range of full N-body simulations, without any need for fine tuning of the model parameters. We introduce an ansatz that dynamical friction ceases when $b_{\rm min} > b_{\rm max}$, as there are no valid impact parameters for encounters that would contribute to the frictional force (see section section \ref{R_2006.ch}). We describe and test a method of applying equation \ref{dynfric.eq} to a cluster of particles, and implement this model in the direct summation code NBODY6 \citep{b11}. This opens up a wide array of astrophysically interesting problems, for example following the internal dynamics of collisional star clusters while simultaneously capturing their orbital decay due to dynamical friction in the inner galaxy.

The paper is organised as follows. In section \S\ref{theory.ch} we describe the physical motivation for our choice of impact parameters and the treatment of $b\sscript{max}$ in cored profiles. Section \S\ref{implementation.ch} describes how Chandrasekhar's formula is applied to satellites comprised of a cluster of particles and its implementation in NBODY6. Section \S \ref{IC.ch} describes the galaxy and cluster models used throughout the paper. In section \S\ref{results.ch} we compare the semi-analytic results of our modified version of NBODY6 (hereafter NBODY6df) with full $N$-body models computed in NBODY6 and the tree-code GADGET \citep{b9}. Finally we compare with previous work in section \S\ref{discussion.ch}, and conclude our results in section \S\ref{conclusion.ch}.

\section[]{Theory}
\label{theory.ch}

Chandrasekhar's dynamical friction formula for a satellite of mass $M\sscript{S}$ is often written as \citep{b2}:

\begin{equation}
	\frac{d\bmath{v\sscript{S}}}{dt} = -4\pi G^2 M\sscript{S} \rho \log(\Lambda)f(v\sscript{*})\frac{\bmath{v\sscript{S}}}{v\sscript{S}^3},
	\label{dynfric.eq}
\end{equation}
where $\log(\Lambda)$ is the Coulomb logarithm equal to $ \log{(b\sscript{max}/b\sscript{min})}$, $\bmath{v\sscript{S}}$ is the satellite velocity, $v\sscript{S} = \bmath{|v\sscript{S}|}$ and $\rho$ is the local background density. If we assume a Maxwellian distribution of velocities:

\begin{equation}
	f(v\sscript{*}) = \mathrm{erf} (X) - \frac{2X}{\sqrt{\pi}}\exp(-X^2),
	\label{fvstar.eqn}
\end{equation}
where $X = v\sscript{S}/\sqrt{2}\sigma$.

Equation \ref{dynfric.eq} is derived under the assumption of an infinite homogeneous background. Despite this assumption, the Coulomb logarithm, $\log(\Lambda)$, takes into account the finite size of a real system through the maximum and minimum impact parameters. Typically $b\sscript{max} $ is of the order of the size of the host system, and $b\sscript{min} $ is defined as the impact parameter for a 90 degree deflection.

From a theoretical standpoint, it is surprising that Chandrasekhar's formula is so successful at reproducing the effects of dynamical friction. In real systems, dynamical friction almost certainly results from discrete resonances with background stars \citep{b27,b28}. Chandrasekhar's formula does not model these resonances, however it likely works because in most systems when an infalling body migrates from one radius to the next it experiences a whole new set of resonances. \citet{b27} show that if the resonances are assumed to form a continuum, Chandrasekhar's approximation is reproduced. In this way, the infalling body behaves similarly to a massive body moving through an infinite homogeneous medium that encounters each background star only once. If we construct a special system, however, where by moving from one radius to the next we do not encounter new resonances then Chandrasekhar's formula has been known to fail. An example of this is given by the harmonic core (where the background density is constant: $\rho(r) \sim \rho\sscript{0}$), which exhibits a short burst of super-Chandrasekhar dynamical friction followed by rapid stalling of the cluster orbit \citep{b24,b39,b35}.

\subsection{Maximum impact parameter}

Although a constant $b\sscript{max}$ has been traditionally used to estimate inspiral timescales, \citet{b12} computed semi-analytic approximations of $N$-body models and found that a spatially dependent $b\sscript{max}$ better reproduces simulation results. The physical motivation for this comes from the local approximation under which equation \ref{dynfric.eq} is derived, which assumes the density distribution is constant up to $b\sscript{max}$. Therefore $b\sscript{max}$ should be a local property of the system. The authors took $b\sscript{max}$ to be the distance from the Galactic Centre, $R\sscript{g}$, which seems like a reasonable order of magnitude estimate, as  the density of particles with impact parameters larger than $R\sscript{g}$ is low compared to the local density distribution around the subject cluster. However for sufficiently cuspy profiles ($\gamma > 1$, where $\gamma$ is the asymptotic slope of the distribution) this approach will typically over estimate the dynamical friction effect near the centre of the background distribution. The slope of the density distribution is difficult to account for, however \citet{b22} show that:

\begin{equation}
		b\sscript{max} = \frac{\rho(R\sscript{g})}{\left.\frac{d\rho}{dr}\right|\sscript{R\sscript{g}}},
	\label{bmax0.eq}
\end{equation}
is a better maximum cutoff to compensate for the cuspy density profile (i.e. the local density over the local density gradient). This impact parameter gives a length scale for which the density is approximately constant, giving a more accurate representation of the local approximation \citep[see also][]{b21}. This makes intuitive sense if one considers the two extreme cases of density distribution. If $\rho$ is a constant over all space, $b\sscript{max} \rightarrow \infty$, and the force logarithmically diverges, as in Chandrasekhar's original derivation. On the other hand if the distribution is infinitely cuspy (i.e. $\rho(r)\sim r^{-\infty}$), $b\sscript{max} \rightarrow 0$, and the satellite effectively orbits a point mass in a Keplerian orbit with no decay.

From equation \ref{bmax0.eq}, \citet{b22} approximate that $b\sscript{max}\sim R\sscript{g}/\gamma$, however we keep the full expression in this paper so that the denominator can vary locally. We assume that the background density distribution is given by a Dehnen model \citep{b13}. For this model the density and its derivative are both analytic, and Equation \ref{bmax0.eq} can be expressed as:
 
 \begin{equation}
	b\sscript{max} =  \frac{R\sscript{g}(R\sscript{g} + a)}{a\gamma + 4R\sscript{g}},
	\label{bmax0_dehnen.eq}
\end{equation}
which indeed reduces to $b\sscript{max}=R\sscript{g}/\gamma$ for $R\sscript{g}<<a$. Note that an attractive feature of equation \ref{bmax0_dehnen.eq} is that it is well-behaved in the limit $\gamma \rightarrow 0$, tending to a constant $b\sscript{max}=a/4$. This is not the case if we use instead $b\sscript{max}= R\sscript{g}/\gamma$ for which $b\sscript{max} \rightarrow \infty$. This led \citet{b22} to adopt $b\sscript{R\sscript{g}} = R\sscript{g}$ for $\gamma<1$. It would seem, then, that our equation \ref{bmax0_dehnen.eq} is superior in this regard. However, the finite $\gamma \rightarrow 0$ limit is peculiar to the assumed background Dehnen profile. It is straightforward to show that split power law profiles that transition from the inner to the outer slope more steeply than the Dehnen profile will a produce divergent $b\sscript{max}$ in the limit $\gamma \rightarrow 0$, if we assume the ansatz $b\sscript{max} = \rho(R\sscript{g})/\Delta\rho(R\sscript{g})$. For this reason, in this paper we follow \citet{b22} and assume $b\sscript{max} = R\sscript{g}$ for $\gamma < 1$.

\subsection{Minimum Impact Parameter}

The minimum impact parameter (i.e. the impact parameter corresponding to a 90 degree deflection) of extended objects is roughly of the order of the half mass radius of the object \citep{b2}. \citet{b12} found, for infalling satellites of a Plummer density profile, that $b_{min}$ is well approximated by $1.4 \epsilon_{s}$, where $\epsilon_{s}$ is the Plummer scale radius, $a$. In terms of the half-mass radius this corresponds to $b\sscript{min} = (1.4/1.3) r\sscript{hm}$.

It should be noted that even though \citet{b12}'s $b\sscript{min}$ was fit for a Plummer sphere, it is a reasonable approximation for other stellar distributions. \citet{b21} show in their discussion about $b\sscript{min}$ that the minimum impact parameter for Plummer, King and singular isothermal sphere models are very similar in terms of the half mass radius. Similarly to \citet{b21}, we take $b\sscript{min}$ of extended to objects to be $r\sscript{hm}$ instead of \citet{b12}'s $\sim1.07r\sscript{hm}$, to keep our formalism physically motivated rather than calibrated by $N$-body models.

\begin{figure}
 \includegraphics[width=\linewidth]{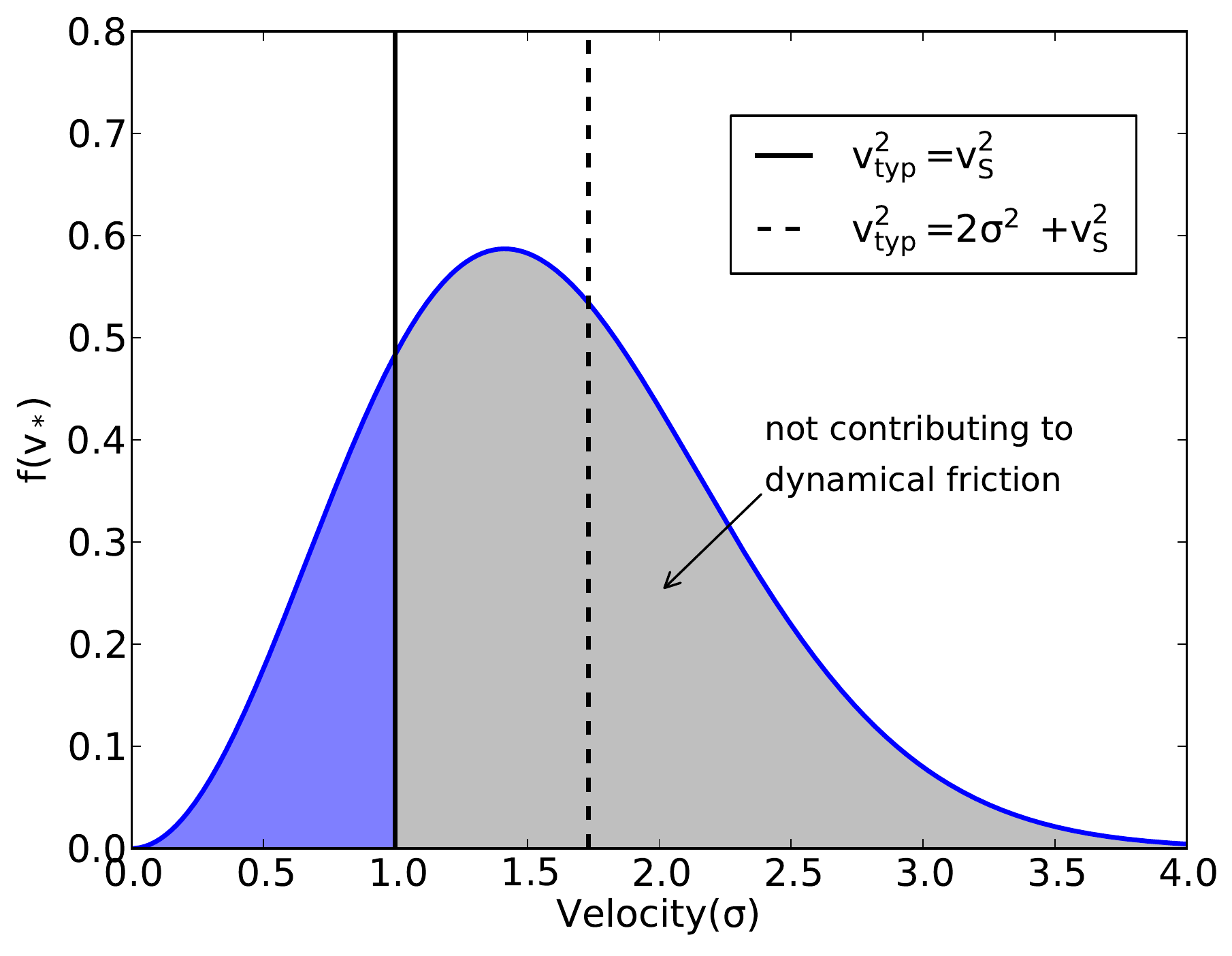}
 \caption{Schematic of the velocity distribution function. If $v\sscript{S}\lesssim\sigma$, then \citet{b22}'s $v\sscript{typ}$ significantly overestimates the maximum impact velocity, as, for isotropic distributions, background stars moving faster than the satellite (the shaded area) are not expected to induce dynamical friction. The schematic assumes a Maxwellian velocity distribution and is normalised to units of $\sigma$.} 
 \label{b90.fig}
\end{figure}

At any epoch, we take $b\sscript{min}$ to be the maximum of  $r\sscript{hm}$ and the minimum impact parameter of a point mass, which is typically taken to be $GM/v\sscript{typ}^2$ (where $M$ and $v\sscript{typ}$ are the bound mass and "typical" velocity for an encounter, respectively) \citep{b2}. In general, $v\sscript{typ}$ (and thus $b\sscript{min}$) are poorly constrained. \citet{b21} take $v\sscript{typ}^2 = 2\sigma^2 + v\sscript{S}^2$. If one considers that $b$ will be minimised for the highest velocity encounter, this seems a reasonable choice at first glance. However, with this formulation $v\sscript{typ}$ is the maximum relative velocity of \textit{any} encounter (i.e. the tail of the Maxwellian velocity distribution), not the maximum velocity of encounters that actually contribute to dynamical friction. With this prescription when $\sigma^2 \sim v\sscript{S}^2$ (which can be the case at the apocentre of eccentric orbits, or throughout the entire core in shallow profiles) the minimum impact parameter will be largely underestimated. This is because the relative velocity of the encounter is larger than the velocity of the satellite, and (for isotropic background distributions) only particles moving slower than the satellite are assumed to contribute to dynamical friction \citep{b2}. The Maxwellian velocity distribution function already selects only the stars that have velocities slower than the satellite to contribute to the friction, and as such only the relative velocities of these stars need to be taken into account, for consistency. For these reasons, we propose that $v\sscript{typ}$ should really be interpreted as $v\sscript{S}$, the velocity of the infalling body. This is the maximum velocity encounter that can contribute to dynamical friction and will therefore give rise to the smallest impact parameter. As such we redefine $v\sscript{typ}^2 = v\sscript{S}^2$. Fig. \ref{b90.fig} displays this schematically. The implications of using this corrected minimum impact parameter over \citet{b22}'s prescription are discussed in section \ref{gt_shallow.ch}.

\subsection{The Coulomb logarithm and core stalling}

These prescriptions for $b\sscript{max}$ and $b\sscript{min}$ give us the following functional form for the Coulomb logarithm:

\begin{equation}
	\log(\Lambda) = \log\left(\frac{b\sscript{max}}{b\sscript{min}}\right) = \log\left(\frac{\mathrm{min}(\rho(R\sscript{g})/\Delta\rho(R\sscript{g}),R\sscript{g})}{\mathrm{max}\left(r\sscript{hm}, GM\sscript{S}/v\sscript{S}^2\right)}\right).
	\label{coulog.eq}
\end{equation}
Equation \ref{coulog.eq} shows that our prescription for $\log\Lambda$ is a function of the radial distance to the centre of the background potential, the slope of the background distribution and the half mass radius of the cluster.  If during inspiral $b\sscript{min} \geq b\sscript{max}$, the dynamical friction term is set to zero. This ansatz is reasonable since this means there are no particles available to scatter off the satellite in a way that would reduce its orbital energy. We now show that this ansatz is equivalent to the well known result that friction ceases if the satellite mass approaches the enclosed mass of the background \citep{b2}, as: 
\begin{subequations}
\begin{align}
v\sscript{typ}^2 &\sim \frac{GM\sscript{g}(R\sscript{g})}{R\sscript{g}} \sim \frac{GM\sscript{g}(R\sscript{g})}{b\sscript{max}},\\
b\sscript{min}&\sim \frac{GM}{v\sscript{typ}^2} \sim \frac{M}{M\sscript{g}(R\sscript{g})}b\sscript{max},\\
\frac{b\sscript{min}}{b\sscript{max}} &\sim \frac{M}{M\sscript{g}(R\sscript{g})},
\end{align}
\label{stalling.eq}
\end{subequations}
where $M\sscript{g}(R\sscript{g})$ is the galaxy mass enclosed at $R\sscript{g}$.

Stalling occurs at this scale because perturbations from individual stars dominate over the mean field effects, making dynamical friction less efficient \citep{b26}.

For the case of a Dehnen model an approximate analytic equation for the stalling radius can be derived. Equating the argument of the  Coulomb logarithm to unity and assuming a circular orbit:

\begin{subequations}
\begin{align}
\frac{b\sscript{max}}{b\sscript{min}} &= \frac{R\sscript{g}(R\sscript{g} + a)/4R\sscript{g}}{GM\sscript{S}/v\sscript{S}^2},\\
v^2\sscript{S} &= \frac{GM\sscript{g}(R\sscript{g})}{R\sscript{g}},\\
\frac{b\sscript{max}}{b\sscript{min}} &= \frac{R\sscript{g}(R\sscript{g} + a)/4R\sscript{g}}{M\sscript{S}R\sscript{g}/M\sscript{g}(R\sscript{g})},\\
\label{rsdev_d}
\frac{b\sscript{max}}{b\sscript{min}} &= \left(\frac{R\sscript{g} + a}{4R\sscript{g}}\right)\left(\frac{M\sscript{g}(R\sscript{g})}{M\sscript{S}}\right) = 1.
\intertext{Recalling the formula for $M\sscript{g}(R\sscript{g})$ \citep{b13}:}
M\sscript{g}(R\sscript{g}) & = M\sscript{g}\left(\frac{R\sscript{g}}{R\sscript{g} + a}\right)^{3-\gamma},\\
\intertext{and inserting this into equation \ref{rsdev_d} and rearranging:}
\frac{M\sscript{S}}{M\sscript{g}} &=\left(\frac{R\sscript{g}}{R\sscript{g}+a}\right)^{3-\gamma}\frac{R\sscript{g} + a}{4R\sscript{g}},\\
\label{rsdev_g}
\frac{M\sscript{S}}{M\sscript{g}} &= \frac{r^{3-\gamma}}{(r+a)^{2-\gamma}(a\gamma + 4r)}.
\intertext{If we take the limit of $r<a$:}
\frac{M\sscript{S}}{M\sscript{g}} &= \frac{r^{3-\gamma}}{a^{2-\gamma}+a\gamma}.\\
\intertext{Therefore:}
R\sscript{S} &= \left(\frac{M\sscript{S}}{M\sscript{G}} (a^{2-\gamma}+a\gamma)  \right)^{\frac{1}{3-\gamma}},
\end{align}
\label{rsdev.eq}
\end{subequations}
where $R\sscript{S}$ is the stalling radius of the satellite. Note that this is the theoretical stalling radius for a point particle. If the particle loses mass, $M\sscript{S}/M\sscript{G}$ will shrink and the cluster can potentially reach further in, but of course the timescale for inspiral will be longer. We will show in section \S\ref{gt_shallow.ch} that this shrinking $\log(\Lambda)$ captures the unique core stalling in shallow profiles when coupled with $f(v\sscript{*})$. For a profile with an intrinsically flat core, stalling occurs even farther out, this phenomenon is explained in section \ref{R_2006.ch}.

\subsection{Velocity Dispersion}

The fraction of background stars moving slower than the satellite (equation \ref{fvstar.eqn}) is obtained from the underlying density distribution. Given a particular analytic density distribution, the velocity dispersion as a function of $R\sscript{g}$ can be derived from the Jeans equation. For a Dehnen model $\sigma(R\sscript{g})$ is analytic if $4\gamma$ is an integer, and in the current implementation of NBODY6df a selection of analytic results have been implemented for various values of gamma. Once derived this allows for a quick analytical calculation in the code (see Appendix for the full derivation). To use a non-integer value of $4\gamma$ one would need to implement a numerical solver of the Jeans equation in the code. However, for the sake of speed, we suggest instead to look for a degenerate model by modifying the scale radius, $a$, and mass, $M\sscript{G}$, of the Dehnen model so that an integer $4\gamma$ may be used. We chose to use Dehnen models in the initial implementation due to their versatility for modelling spherical systems. If one would like to implement a different density distribution, we urge authors to take great care with the definition of $b\sscript{max} = \rho\sscript{0}/\Delta\rho(r)\sscript{0}$, however this impact parameter is calculable for any density distribution with a continuous derivative (see also \citet{b21,b22}).

\section{Implementation}
\label{implementation.ch}

We implemented equation \ref{dynfric.eq} as an external analytical acceleration in NBODY6df. Dynamical friction is applied on the regular integration step, which is computed in parallel on the GPU.

\citet{b14} fit semi-analytical models to $N$-body simulations of dwarf galaxies experiencing dynamical friction in larger parent galaxies. It is shown that the simulations undergo enhanced dynamical friction as compared with Chandrasekhar's formula due to two effects. The first is direct gravitational interactions with escaped particles. This is included in NBODY6df by integrating tidally stripped material self consistently. The second is the indirect effect of material that is energetically unbound but remains close to the cluster, enhancing the gravitational wake in the background medium. In an attempt to replicate this effect, the mass $M\sscript{S}$ in equation \ref{dynfric.eq} is taken to be the total mass of the particles contained inside the Roche volume, i.e. the particles for which $F\sscript{cluster} > F\sscript{tidal}$, where $F\sscript{cluster}$ is the magnitude of the gravitational force on the star due to the $N$-body particles and $F\sscript{tidal}$ is the tidal force experienced by the star. $F\sscript{tidal}$ is defined as $|\bmath{F\sscript{g}}-\bmath{F\sscript{0}}|$, where $\bmath{F\sscript{g}}$ is the force experienced by the star due to the background distribution, and $\bmath{F\sscript{0}}$ is the similar force felt by the density centre of the cluster. This procedure ensures correct calculation of Roche-volume membership and is scale independent, requiring no approximate tidal radius. The force exerted by the background distribution is calculated using existing NBODY6 routines. If the user wishes to change the background potential they can simply replace the potential in NBODY6df in the same manner they would in NBODY6, however we again stress that the velocity dispersion calculation and maximum impact parameter need to be updated too, as discussed in the previous section. We will refer to particles inside the Roche volume as``bound" for ease of use, even though this includes potential escapers. Particles experience no dynamical friction whilst unbound, but will feel dynamical friction once again if they re-enter the Roche volume.
NBODY6 includes the regularisation of binaries and close encounters \citep{b15} in which the system is replaced by a centre of mass (CoM) particle during the regular step. The regularised system is considered under the dynamical friction regime if its CoM particle is bound. The dynamical friction force is then applied directly to the CoM particle, and the differential force on each particle is handled by the KS regularisation scheme.

Collecting all of the scalar terms in equation \ref{dynfric.eq} allows it to be rewritten as:
\begin{equation}
	\bmath{a}\sscript{df} = -C\sscript{fric}(\bmath{v}\sscript{S} /v^3\sscript{S}),
	\label{dynfric_c.eq}
\end{equation}
where:
\begin{equation}
	C\sscript{fric} = 4\pi M\sscript{Roche}\log(\Lambda)\rho(R\sscript{g})\left[\mathrm{erf}(X) - \frac{2X}{\sqrt{\pi}}\exp(-X^2) \right].
	\label{dynfcoef.eq}
\end{equation}
The cluster velocity, $\bmath{v}\sscript{S}$, is taken to be the average velocity of the particles in the cluster core with respect to the galactocentric rest frame. Equation \ref{dynfcoef.eq} is calculated every time NBODY6df adjusts important parameters. This coefficient is used in all dynamical friction calculations until the next adjust routine. The coefficient varies slowly and the computed value is approximately constant between parameter adjustments so long as the adjust time is significantly lower than the orbital period.

When the dynamical friction correction is applied to a particle, the change in energy is calculated and added to a variable representing the total energy removed from the system due to dynamical friction. When NBODY6df calculates the total energy of the system we include this term so that total energy is conserved, and the energy error in the $N$-body calculation can be evaluated in the usual way.

\section{Simulations}
\label{IC.ch}

\begin{table*}
\begin{minipage}{170mm}
\centering

  \begin{tabular}{@{}llrrrrrrrl@{}}
   Run Name & Code & $N\sscript{S}$ & $m\sscript{*S}$ & $N\sscript{bg}$
     & $m\sscript{*bg}$ & $R\sscript{a}$ & $\gamma$ & Eccentricity & Section\\
   & & & ($\unit{M_{\odot}}$) &
     & ($\unit{M_{\odot}}$) & ($\unit{pc}$) \\
 \hline\hline
 df1k & NBODY6df & 1k & 100.0 & - & - & 10.0 & 1.5 & 0.0 & \ref{nb6_discussion.ch}\\
 df1ke & NBODY6df & 1k & 100.0 & - & -& 10.0 & 1.5 & 0.3\\
 df1ke0.75 & NBODY6df & 1k & 100.0 & - & - & 10.0 & 1.5 & 0.75\\
 df1kg1 & NBODY6df & 1k & 100.0 & - & - & 10.0 & 1.0 & 0.0\\
 df1kg1.75 & NBODY6df & 1k & 100.0 & - & - & 10.0 & 1.75 & 0.0\\
 nb1k & NBODY6 & 1k & 100.0 & 464k & 100.0 & 10.0 & 1.5 & 0.0\\
 nb1ke & NBODY6 & 1k & 100.0 & 464k & 100.0 & 10.0 & 1.5 & 0.3\\
 nb1ke0.75 & NBODY6 & 1k & 100.0 & 464k & 100.0 & 10.0 & 1.5 & 0.75\\
 nb1kg1 & NBODY6 & 1k & 100.0 & 429k & 100.0 & 10.0 & 1.0 & 0.0\\
 nb1kg1.75 & NBODY6 & 1k & 100.0 & 483k & 100.0 & 10.0 & 1.75 & 0.0\\ 
 \hline
  dfL5 & NBODY6df & 1k & 100.0 & - & - & 5.0 & 1.5 & 0.0 & \ref{nb6_angmom.ch}\\
 dfL5e & NBODY6df & 1k & 100.0 & - & - & 5.0 & 1.5 & 0.3\\
 dfL5.0e0.75 & NBODT6df & 1k & 100.0 & - & - & 5.0 & 1.5 & 0.75\\
 dfL2.5 & NBODY6df & 1k & 100.0 & - & - & 2.5 & 1.5 & 0.0\\
 dfL2.5e & NBODY6df & 1k & 100.0 & - & - & 2.5 & 1.5 & 0.3\\
 dfL2.5e0.75 & NBODY6df & 1k & 100.0 & - & - & 2.5 & 1.5 & 0.75\\
 nbL5 & NBODY6 & 1k & 100.0 & 464k & 100.0 & 5.0 & 1.5 & 0.0\\
 nbL5e & NBODY6 & 1k & 100.0 & 464k & 100.0 & 5.0 & 1.5 & 0.3\\
 nbL5.0e0.75 & NBODY6 & 1k & 100.0 & 464k & 1000.0 & 5.0 & 1.5 & 0.75\\
 nbL2.5 & NBODY6 & 1k & 100.0 & 464k & 100.0 & 2.5 & 1.5 & 0.0\\
 nbL2.5e & NBODY6 & 1k & 100.0 & 464k & 100.0 & 2.5 & 1.5 & 0.3\\
 nbL2.5e0.75 & NBODY6 & 1k & 100.0 & 464k & 1000.0 & 2.5 & 1.5 & 0.75\\
 dfL2k & NBODY6df & 2k & 50.0 & - & - & 5.0 & 1.5 & 0.0\\
 dfL2ke & NBODY6df & 2k & 50.0 & - & - & 5.0 & 1.5 & 0.75\\
 dfL4k & NBODY6df & 4k & 25.0 & - & - & 5.0 & 1.5 & 0.0\\
 dfL4ke & NBODY6df & 4k & 25.0 & - & - & 5.0 & 1.5 & 0.75\\
 dfL8k & NBODY6df & 8k & 12.5 & - & - & 5.0 & 1.5 & 0.0\\
 dfL8ke & NBODY6df & 8k & 12.5 & - & - & 5.0 & 1.5 & 0.75\\  
 nbL2k & NBODY6 & 2k & 50.0 & 464k & 100.0 & 5.0 & 1.5 & 0.0\\
 nbL2ke & NBODY6 & 2k & 50.0 & 464k & 100.0 & 5.0 & 1.5 & 0.0\\
 nbL4k & NBODY6 & 4k & 25.0 & 464k & 100.0 & 5.0 & 1.5 & 0.0\\
 nbL4ke & NBODY6 & 4k & 25.0 & 464k & 100.0 & 5.0 & 1.5 & 0.75\\
 nbL8k & NBODY6 & 8k & 12.5 & 464k & 100.0 & 5.0 & 1.5 & 0.0\\
 nbL8ke & NBODY6 & 8k & 12.5 & 464k & 100.0 & 5.0 & 1.5 & 0.75\\
 \hline
 dfa10 & NBODY6df & 10k & 10.0 & - & - & 10.0 & 1.5 & 0.0 & \ref{gt_discussion.ch}\\
 dfa10e & NBODY6df & 10k & 10.0 & - & - & 10.0 & 1.5 & 0.3\\
 dfa5 & NBODY6df & 10k & 10.0 & - & - & 5.0 & 1.5 & 0.0\\
 dfa2.5 & NBODY6df & 10k & 10.0 & - & - & 2.5 & 1.5 & 0.0\\
 gta10 & GADGET & 10k & 10.0 & 1549k & 30.0 & 10.0 & 1.5 & 0.0\\
 gta10e & GADGET & 10k & 10.0 & 1549k & 30.0 & 10.0 & 1.5 & 0.3\\	
 gta5 & GADGET & 10k & 10.0 & 1549k & 30.0 & 5.0 & 1.5 & 0.0\\
 gta2.5 & GADGET & 10k & 10.0 & 1549k & 30.0 & 2.5 & 1.5 & 0.0\\
 \hline
 gtag0.0 & GADGET & 10k & 10.0 & 1549k & 30.0 & 5.0 & 0.0 & 0.0 &
 \ref{gt_shallow.ch}\\
 gtag0.5 & GADGET & 10k & 10.0 & 1549k & 30.0 & 5.0 & 0.5 & 0.0\\  
 dfg0.0 & NBODY6df & 10k & 10.0 & - & - & 5.0 & 0.0 & 0.0\\
 dfg0.5 & NBODY6df & 10k & 10.0 & - & - & 5.0 & 0.5 & 0.0\\ 
 \hline
 df2k & NBODY6df & 2k & 50.0 & - & - & 10.0 & 1.5 & 0.0 & \ref{nb6_Ndependence.ch}\\
 df4k & NBODY6df & 4k & 25.0 & - & - & 10.0 & 1.5 & 0.0\\
 df8k & NBODY6df & 8k & 12.5 & - & - & 10.0 & 1.5 & 0.0\\
 df16k & NBODY6df & 16k & 6.25 & - & - & 10.0 & 1.5 & 0.0\\
 df32k & NBODY6df & 32k & 3.125 & - & - & 10.0 & 1.5 & 0.0\\
 df64k & NBODY6df & 64k & 1.5625 & - & - & 10.0 & 1.5 & 0.0\\ 
	
\hline\hline

\end{tabular}
\caption{Initial conditions of simulations. Column 1 lists the names of the simulations, where the prefixes: df, nb and gt indicate the code used, NBODY6df, NBODY6 and GADGET, respectively. This is also stated in column 2. Columns 3 to 6 display the particle numbers and masses for both the cluster and the background, subscripts $c$ and $bg$ respectively. Column 7 lists the initial distance of the cluster from the Galactic Centre, all runs start at apocentre. Column 8 states the asymptotic slope used in the background Dehnen model. Column 9 shows the initial eccentricity of the cluster. Column 10 displays which chapter each group of simulations first appears in.}
\label{initial_conditions.tbl}
\end{minipage}

\end{table*}

\subsection{Initial Conditions}
\subsubsection{Cluster}

The clusters in this study are initially Plummer models of mass $M\sscript{S} = 10^5 \unit{M_\odot}$ and half mass radius $r\sscript{hm} = 0.1\unit{pc}$, similar to the clusters modelled in \citet{b8} and \citet{b7}. The mass of a cluster particle is $m\sscript{S} = M\sscript{S}/N\sscript{S}$, where $N\sscript{S}$ is the number of cluster particles.

\subsubsection{Background}

We adopt single component Dehnen models \citep{b13}, representing the central region of the Galaxy. We use a slope $\gamma = 1.5$, scale radius $a = 8.625\unit{pc}$ and mass $M\sscript{g} = 5.9\times 10^7 \unit{M_\odot}$ to represent the density distribution in the central few tens of parsecs in the Milky Way. This closely represents the observed broken power-law profile obtained by \citet{b16} for the central 10 parsecs of the Galaxy. For runs with different $\gamma$ we use the same parameters as stated above. What these profiles represent is arbitrary, and we simply keep the same mass and scale radius for ease of use.

It should be noted that any time-independent analytical spherical background potential can be included in the code, such as the addition of a central SMBH, and a dark halo component (although for some models the density and velocity dispersion functions may need to be calculated numerically). We have adopted a simple model here to ease comparison of the code with full $N$-body models with low-$N$, in which a SMBH may wander significantly. Choosing a single spherical component for the test simulations also gives more applicability to larger scale simulations. Models where the scale radius and mass of the host are larger (i.e. globular clusters in a dwarf spheroidal) will behave in a similar way, as integration in NBODY6 is performed in scale independent Henon units internally, $G=1=M=R\sscript{v}=-4E$ (where $M$ is the total mass, $R\sscript{v}$ is the virial radius and $E$ is the total energy) \citep{b11}. A more realistic treatment of clusters near the Galactic Centre will be investigated in a scientific context in an upcoming paper.

\subsection{Models}

We compare results of NBODY6df with results from fully self-consistent NBODY6 and GADGET runs, where the background distribution is granular. NBODY6 is a direct-summation collisional code and as such we use equal particle masses for the cluster and background to reduce unrealistic scattering. When using the tree-code GADGET however, we use a smaller mass for the cluster particles, as softening reduces the low-$N$ scattering effects. We use softening parameters of $0.025\unit{pc}$ for the cluster particles and $0.1\unit{pc}$ for the bulge particles in all GADGET simulations.

With the maximum initial cluster distance being $10\unit{pc}$ and the background $R\sscript{hm}$ being $\sim15\unit{pc}$ for $\gamma = 1.5$, the truncation of the background at large radii will have a negligible effect on the dynamical friction experienced by the cluster. Thus for both the GADGET and NBODY6 runs we truncate the Dehnen potential at $50\unit{pc}$. The models are summarised in Table \ref{initial_conditions.tbl}.

\section{Results}
\label{results.ch}

\subsection{Comparison with NBODY6}
\subsubsection{Orbit Comparison}
\label{nb6_discussion.ch}

\begin{figure}
 \includegraphics[width=\linewidth]{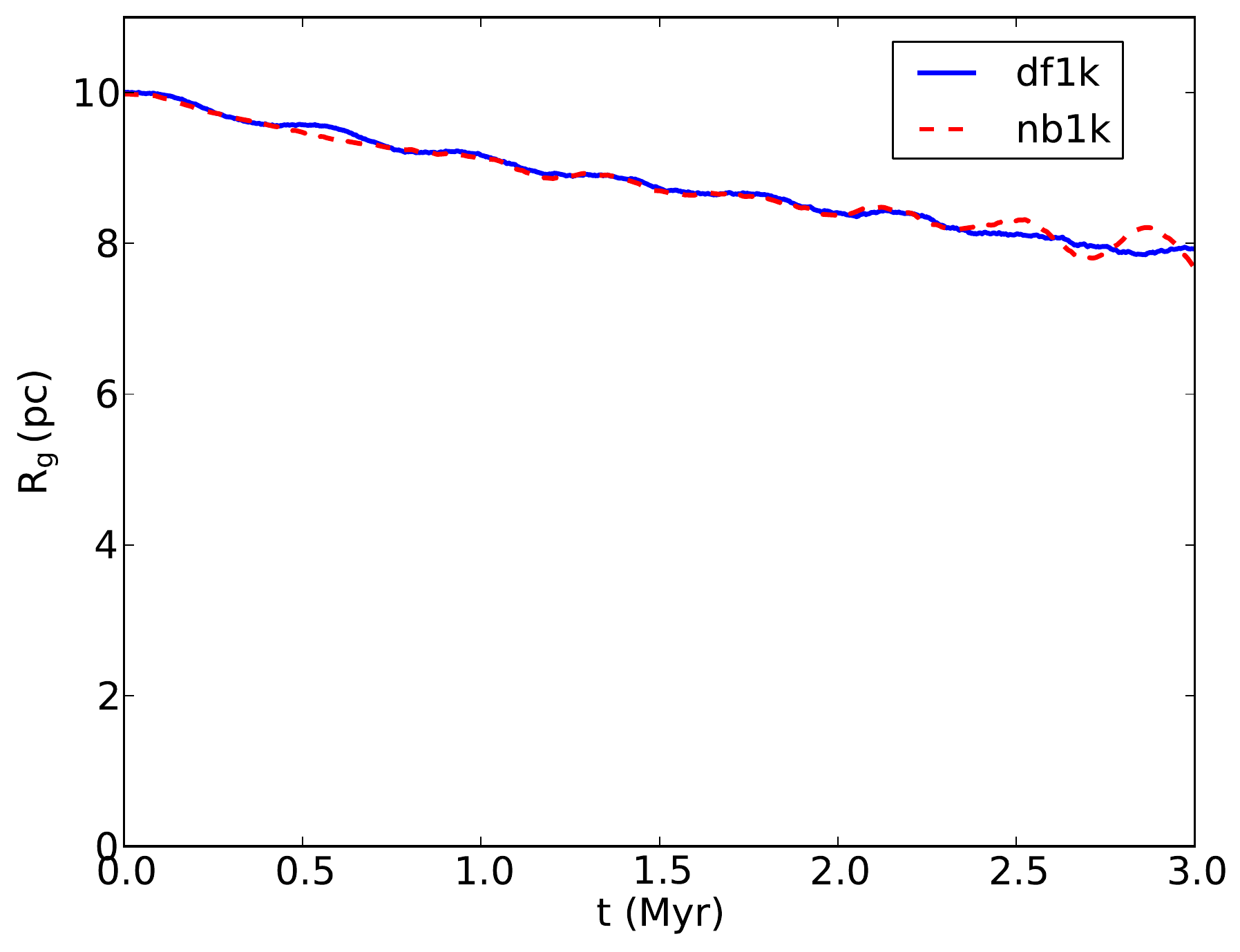}
 \caption{Distance of the cluster with respect to the Galactic Centre as a function of time for df1k (blue line) and nb1k (red dashed line).} 
 \label{1k_Rg.fig}
\end{figure}

\begin{figure}
 \includegraphics[width=\linewidth]{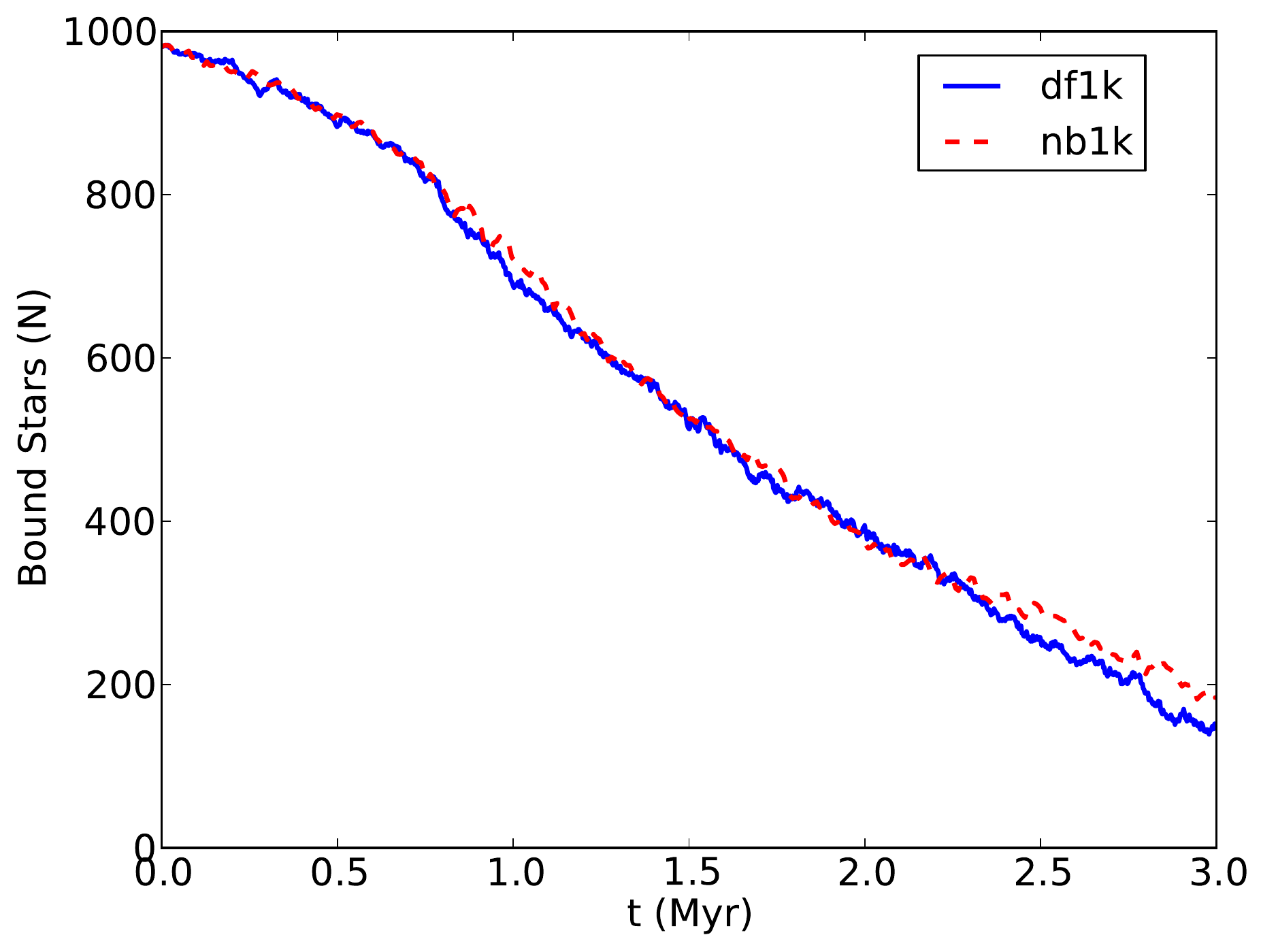}
 \caption{Mass enclosed in the Roche volume as a function of time for df1k (blue line) and nb1k (red dashed line)} 
 \label{1k_bound.fig}
\end{figure}

Simulations df1k and nb1k are compared in Fig. \ref{1k_Rg.fig}, which shows the radial position of the cluster with respect to the Galactic Centre as a function of time. Fig. \ref{1k_bound.fig} shows the bound mass of the clusters in the different simulations. The agreement between the two models is excellent. After $\sim$ 2 Myrs nb1k experiences stochastic changes in its orbit due to the low-$N$ background, this is because the low-$N$ cluster has nearly dissolved by this time. Prior to this epoch, when the clusters are not close to dissolution, the radial distance travelled by the cluster in the two codes differs by less than a few per cent. 

The stochastic changes in the cluster orbit of simulation nb1k come from $N$-body sampling from the distribution function. This introduces chaotic effects on both large and small scales, compared to the equivalent analytic distribution \citep{b17}. At small scales the granularity of the background induces stochastic changes in the orbit if $N$ of the background is low (in nb1k each background particle represents the mass of an ensemble of stars). On large scales the system may deviate from spherical symmetry, inducing moderate eccentricity and precession. These effects accumulate over time and cause the eccentricity in nb1k once the cluster has almost dissolved.

\begin{figure}
 \includegraphics[width=\linewidth]{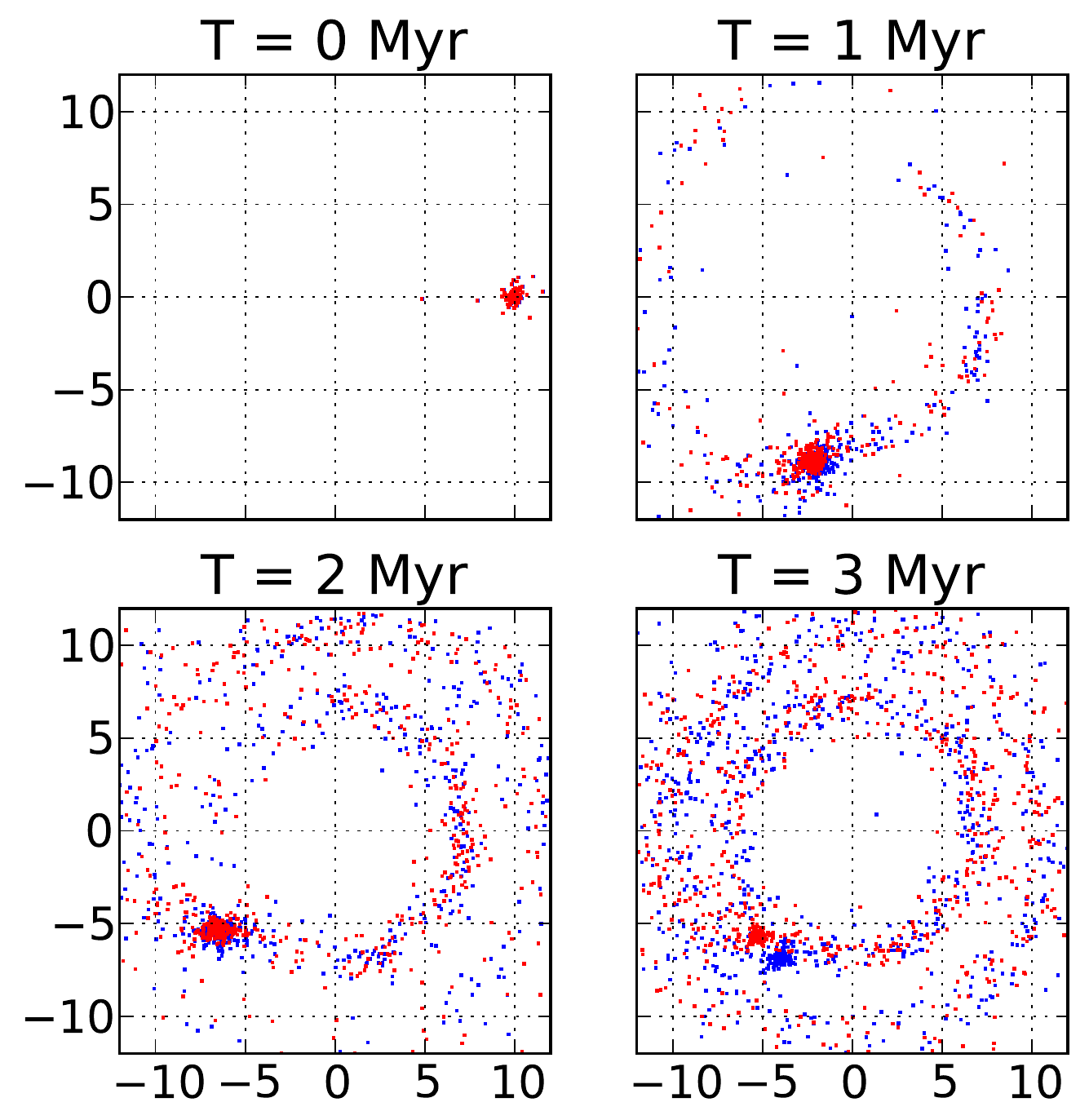}
 \caption{Snapshots of the models df1k (blue) and nb1k (red) at different epochs. For nb1k only particles originating from the cluster are plotted.} 
 \label{1k_snaps.fig}
\end{figure}

Fig. \ref{1k_snaps.fig} shows snapshots of the simulations at different times. Only after $\sim$ 3 Myrs are the models distinguishable, and it can be seen that the structure and distribution of the tidally stripped material is well reproduced in NBODY6df.

\FloatBarrier

\begin{figure}
 \includegraphics[width=\linewidth]{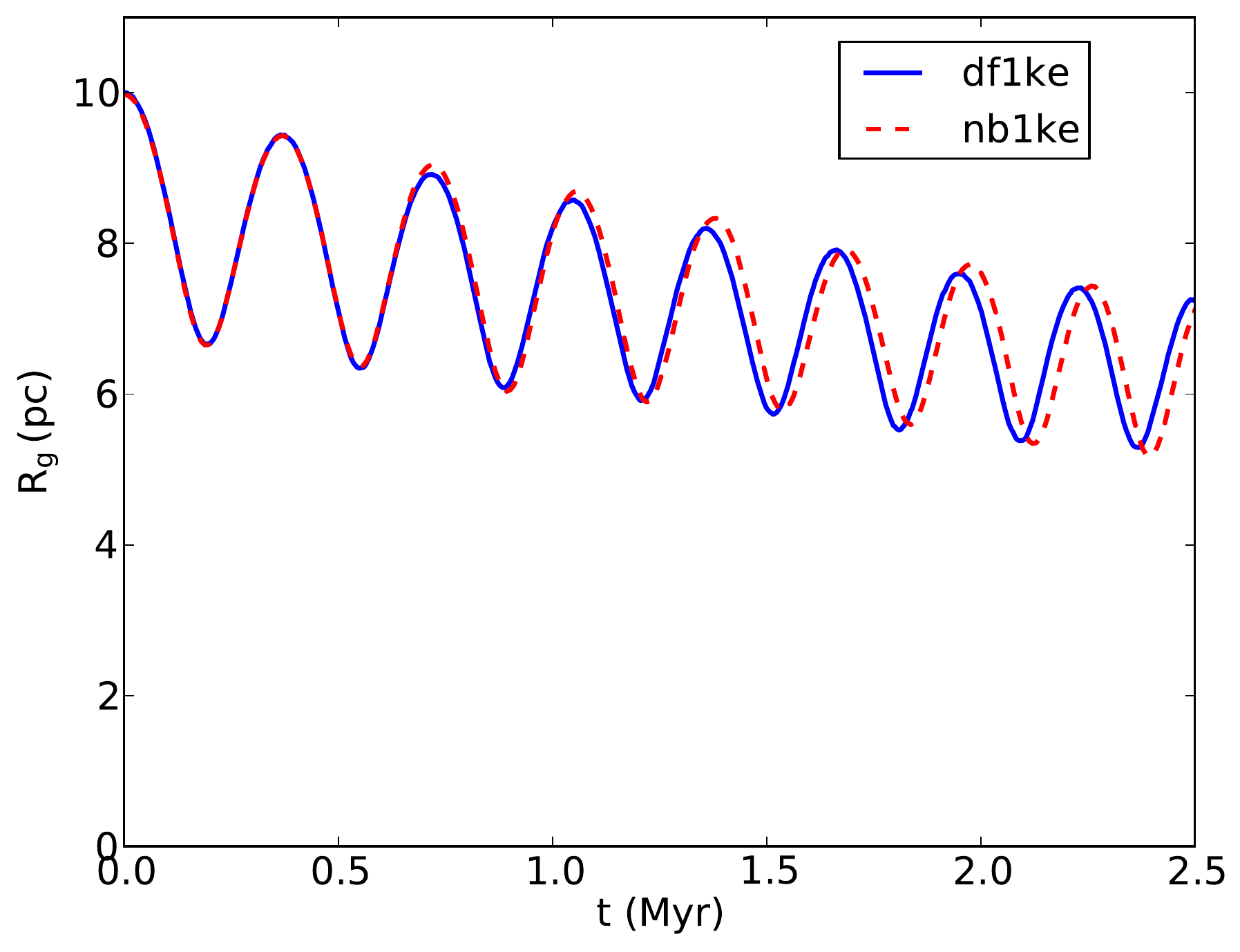}
 \caption{Distance of the cluster with respect to the Galactic Centre as a function of time for df1ke (blue line) and nb1ke (red dashed line).} 
 \label{1ke_Rg.fig}
\end{figure}

\begin{figure}
 \includegraphics[width=\linewidth]{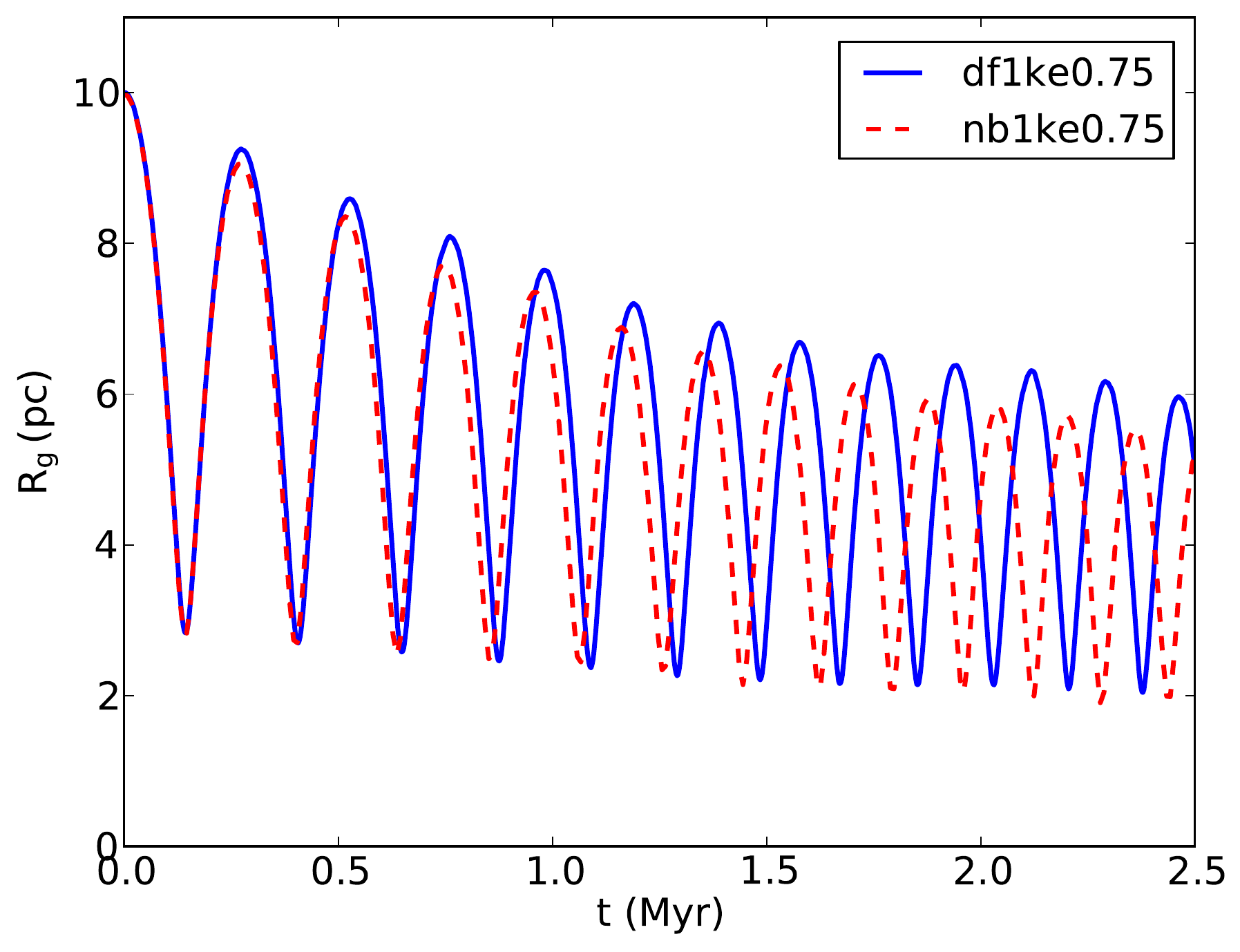}
 \caption{Distance of the cluster with respect to the Galactic Centre as a function of time for df1ke0.75 (blue line) and nb1ke0.75 (red dashed line).} 
 \label{1ke075_Rg.fig}
\end{figure}

Figs. \ref{1ke_Rg.fig} and \ref{1ke075_Rg.fig} show simulations which have the same initial conditions as df1k and nb1k, but with initial cluster eccentricities of 0.3 and 0.75, respectively. The agreement is excellent for both eccentricities. For $e=0.75$ the simulations diverge near the end as the clusters have lost the majority of their mass. The agreement is so good because of our prescription for $b\sscript{min}$, which is dependent on both position and velocity. At pericentre the cluster moves fastest, giving a smaller $b\sscript{min}$ and a stronger dynamical friction force, at apocentre the opposite is true, decreasing the force. Meanwhile $b\sscript{max}$ varies across the length of the orbit due to its radial dependence. The result is an accurate calculation of the force along the entire orbit. The excellent agreement of both these models shows that the semi-analytic dynamical friction scheme in NBODY6df can accurately reproduce the force experienced for a range of eccentricities.

\begin{figure}
 \includegraphics[width=\linewidth]{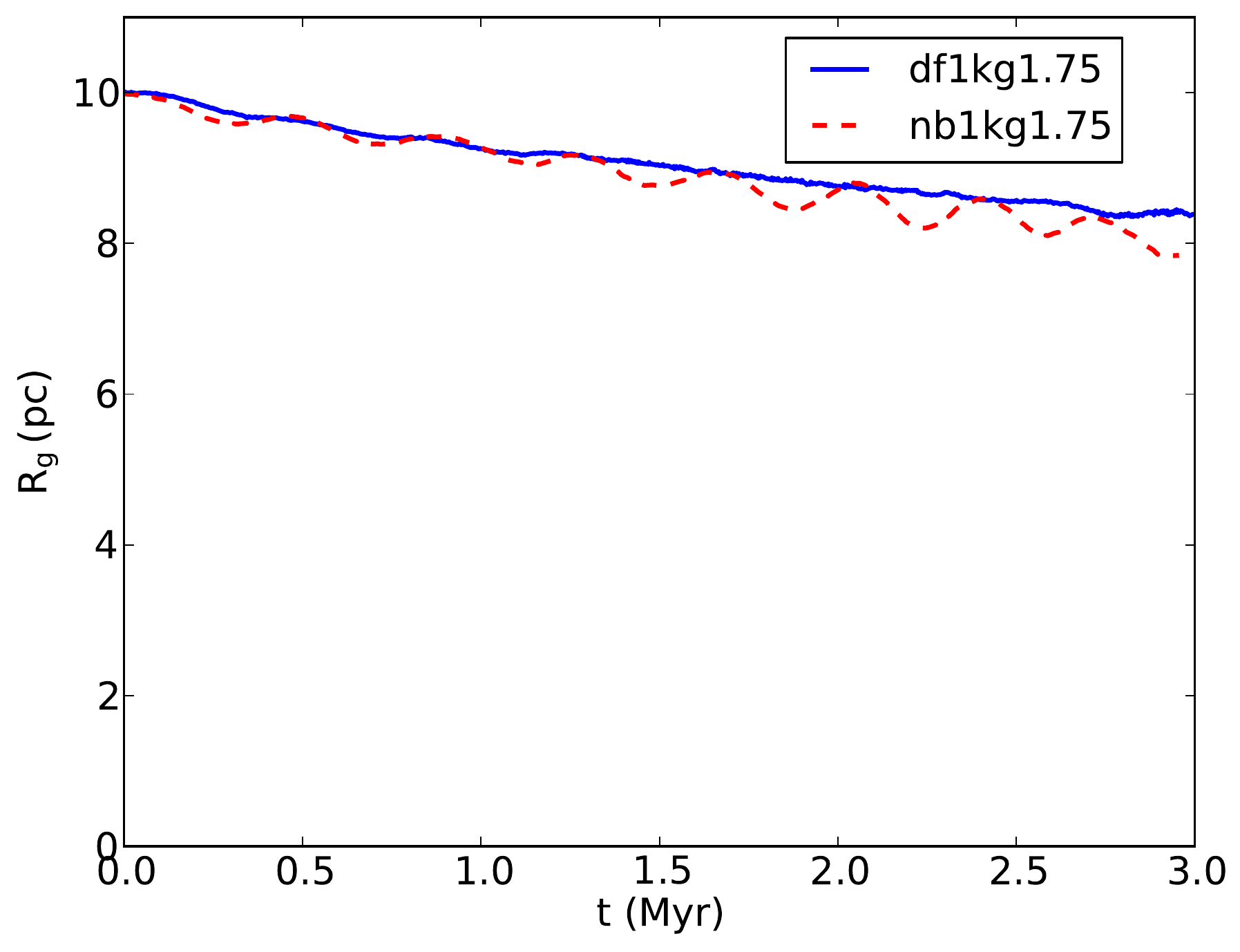}
 \caption{Distance of the cluster with respect to the Galactic Centre as a function of time for df1kg175 (blue line) and nb1kg175 (red dashed line).} 
 \label{1kg175_Rg.fig}
\end{figure}

\begin{figure}
 \includegraphics[width=\linewidth]{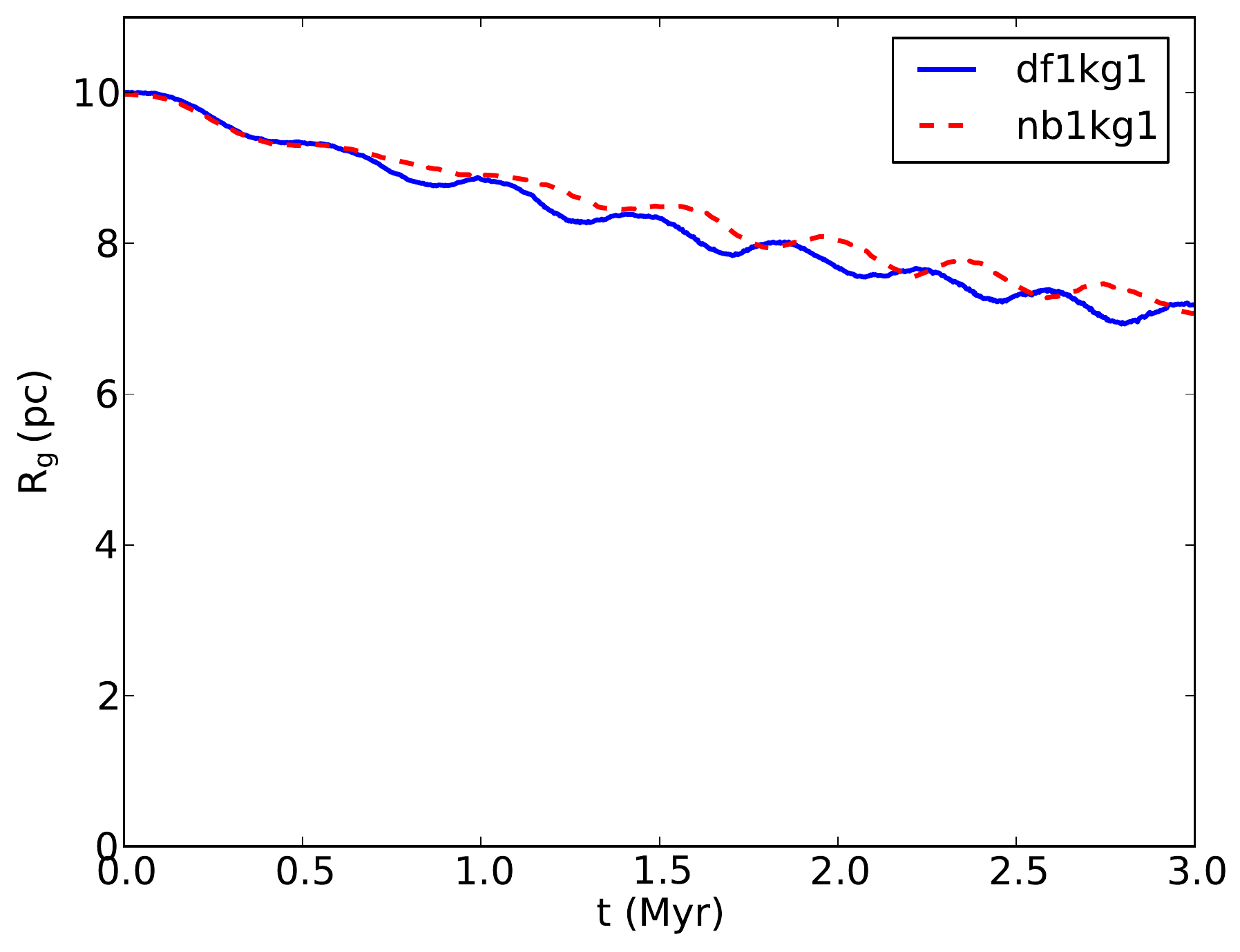}
 \caption{Distance of the cluster with respect to the Galactic Centre as a function of time for df1kg1 (blue line) and nb1kg1 (red dashed line).} 
 \label{1kg1_Rg.fig}
\end{figure}

Figures \ref{1kg175_Rg.fig} and \ref{1kg1_Rg.fig} compare models which have the same initial conditions as df1k and nb1k, but with asymptotic slopes of $\gamma = 1.75$ and $\gamma = 1$ respectively. Both show good agreement. The NBODY6 runs gain some eccentricity from the granularity of the low-$N$ background distributions. This common problem with the low $N$-models is addressed in sections \ref{nb6_angmom.ch} and \ref{gt_discussion.ch}.

\subsubsection{Angular Momentum Comparison}
\label{nb6_angmom.ch}

During inspiral the dynamical friction force is coupled with the relaxation of the cluster, meaning $b\sscript{min}(R\sscript{g},v\sscript{S}) = b\sscript{min}(R\sscript{g},v\sscript{S},t)$ and $M\sscript{cl} = M\sscript{cl}(t)$. Therefore different realisations of low-$N$ simulations can significantly deviate from each other by using a different random seed, as the mass loss from dynamical ejections is very much a stochastic process for low-$N$ simulations, where the relaxation time is short.

Attempting to isolate each effect can give some indication of how accurate an approximation NBODY6df is. In the limit of negligible dynamical friction NBODY6df is identical to NBODY6, and the relaxation timescales will be similar.

\begin{figure}
 \includegraphics[width=\linewidth]{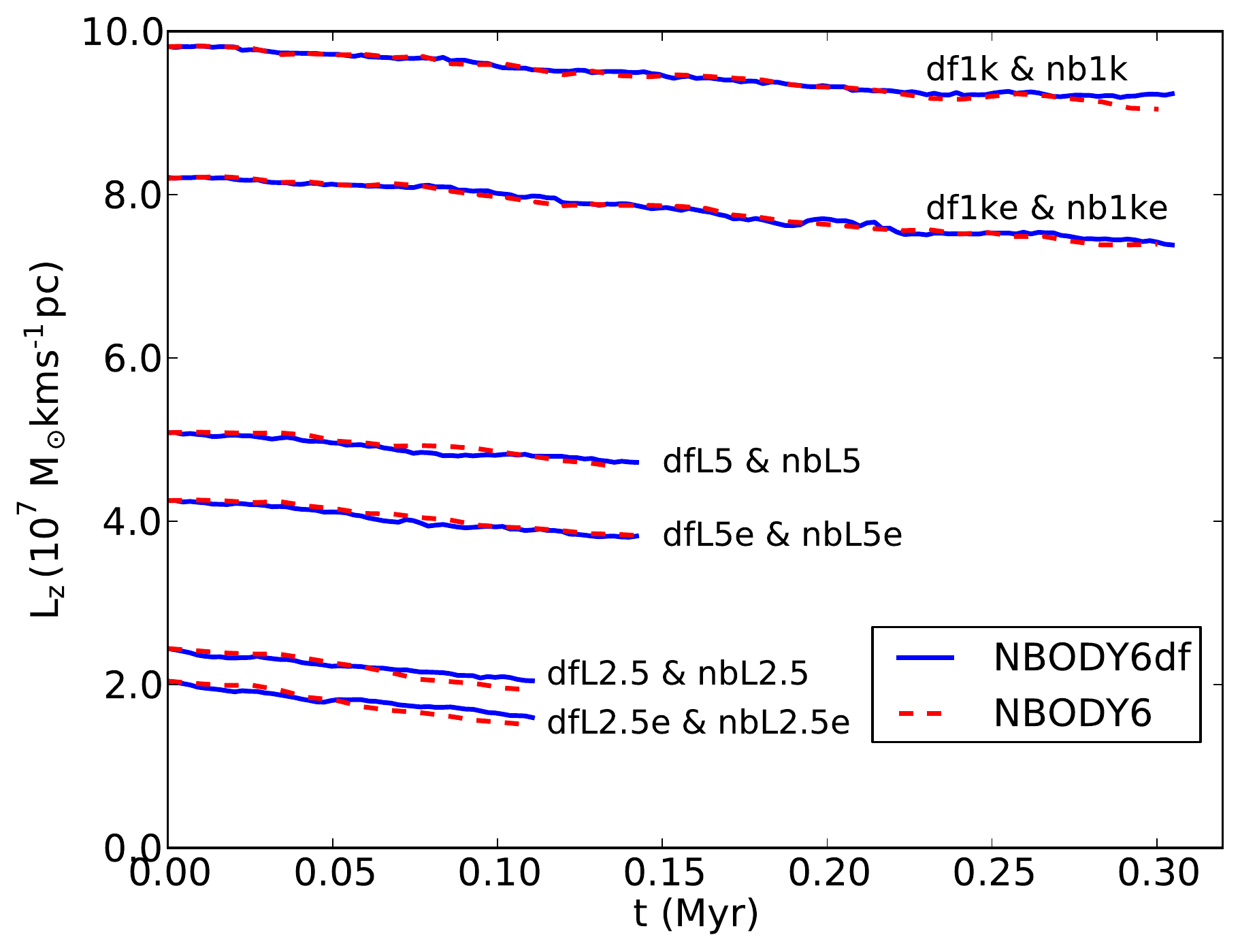}
 \caption{Z-component of angular momentum (perpendicular to the orbital plane) as a function of time for half a cluster orbit at different initial distances 10, 5 and 2.5 pc, and eccentricities of 0.0 and 0.3. In all cases $\gamma = 1.5$.} 
 \label{L_1.5.pdf}
\end{figure}

\begin{figure}
 \includegraphics[width=\linewidth]{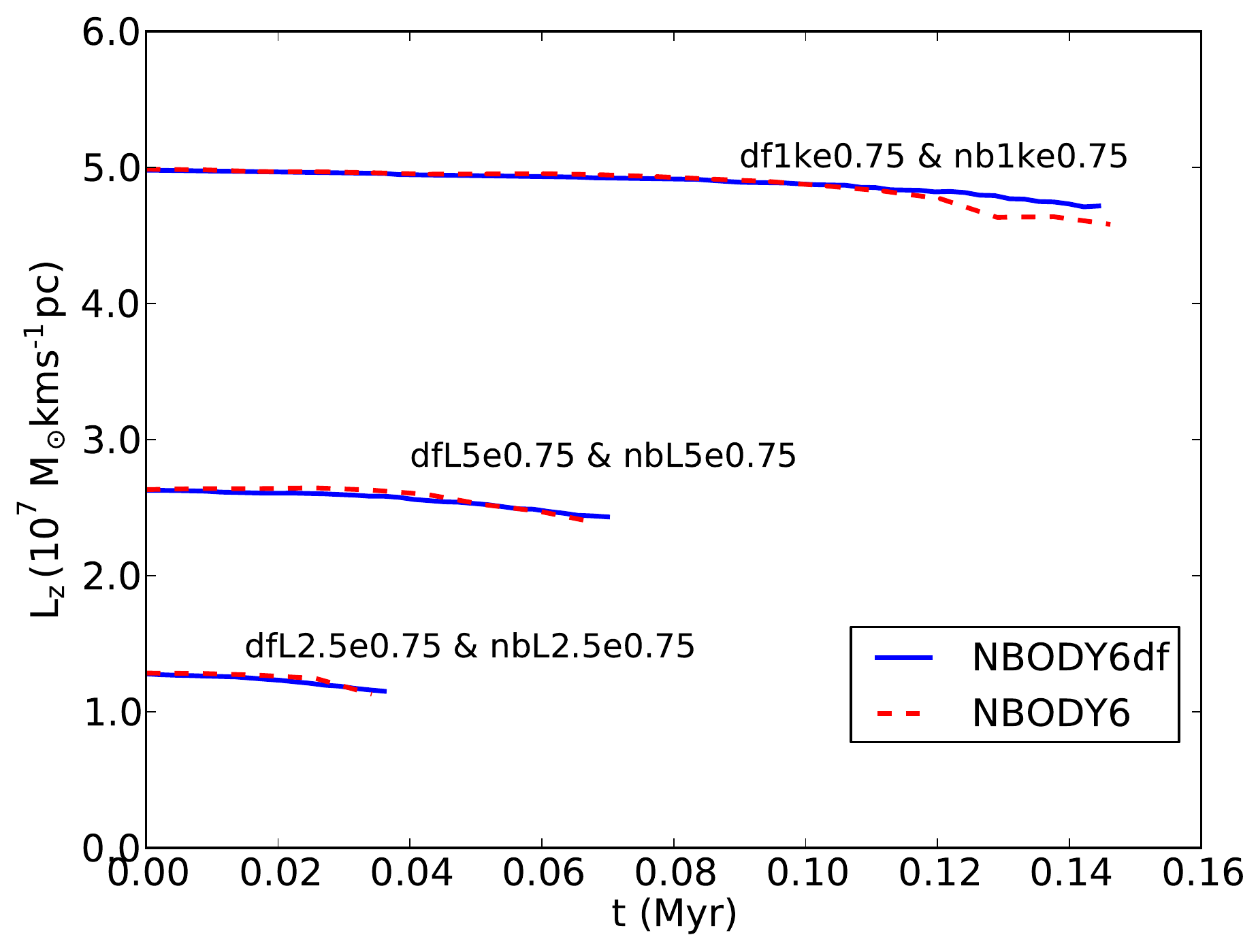}
 \caption{Z-component of angular momentum (perpendicular to the orbital plane) as a function of time for half a cluster orbit at different initial distances 10, 5 and 2.5 pc, with eccentricity of 0.75. In all cases $\gamma = 1.5$.} 
 \label{L_075.pdf}
\end{figure}

We ran a series of short simulations to try to isolate the dynamical friction effect from the relaxation process as much as possible. In Fig. \ref{L_1.5.pdf} we plot the total angular momentum of the bound material perpendicular to the orbital plane (i.e. $L\sscript{z}$) as a function of time for half an orbit, for different initial cluster orbits (see table \ref{initial_conditions.tbl} for initial conditions). Fig. \ref{L_075.pdf} shows the same for eccentricities of 0.75. Over a time of only half an orbit the clusters lose no more than 10 per cent of their mass, and as such the orbital evolution is only weakly dependant on relaxation. In Fig. \ref{L_075.pdf} the full $N$-body models lose a bit more mass than NBODY6df at pericentre. After further investigation this seems to be due to stars near the tidal radius of the cluster being stripped more aggressively due to two body scattering with the low-$N$ background, rather than a deviation from Chandrasekhar's formula.

\begin{figure}
 \includegraphics[width=\linewidth]{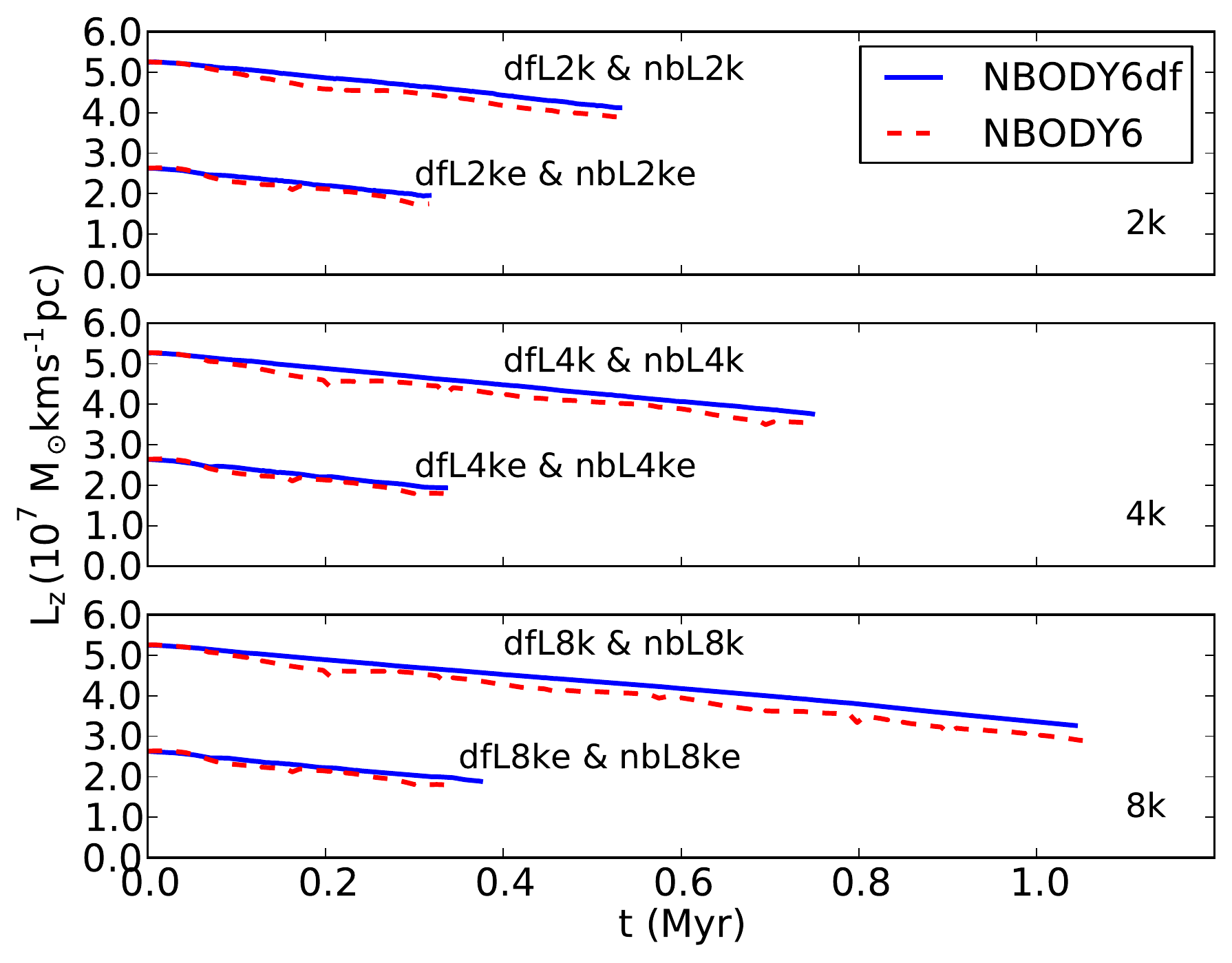}
 \caption{Z-component of angular momentum (perpendicular to the orbital plane) as a function of time until cluster loses 10\% of its mass. At an initial distance of $5\unit{pc}$, witch eccentricities of 0.0 and 0.75. From top to bottom the cluster consists of 2k, 4k and 8k particles. In all cases $\gamma = 1.5$.} 
 \label{L_Ntest.fig}
\end{figure}

We tested a specific case (a=5.0,e=0.0,0.75) in models where the cluster is comprised of 2k, 4k and 8k particles, whilst keeping the total mass constant. We did not redo the entire grid of initial radii and eccentricities as the full $N$-body models are very numerically expensive. The models were run until the cluster in the NBODY6df simulation had lost 10\% of its mass. Fig. \ref{L_Ntest.fig} shows these sets of models, where very good agreement is found. The higher $N$ models retain their mass for longer due to slower two-body relaxation. In all the NBODY6 simulations, the high ratio of cluster particle mass to background particle mass causes a few of the less tightly bound cluster particles to be stripped very early on in the simulation. Therefore the angular momentum of the bound particles is systematically lower after $\sim 0.2 \unit{Myr}$ in all cases. This is not significant, as these stripped particles contribute very little to the total mass of the cluster, and so the gradient of the $L\sscript{z}(t)$ curves still show very good agreement.

The agreement is excellent over the range of orbits tested, which validates that NBODY6df can reproduce the expected angular momentum loss at different radii for both circular and eccentric orbits. The orbital evolution over many orbits can be considered accurate because the dynamical friction coefficient is linear with mass. At any epoch, $t\sscript{1}$, $M\sscript{S} = M\sscript{S}(t\sscript{1})$ and $R\sscript{g} = R\sscript{g}(t\sscript{1})$. If we assume that our limited number of models in Fig. \ref{L_1.5.pdf} indicate that the dynamical friction coefficient is initially correct compared to full $N$-body models at any $M\sscript{S,0},R\sscript{g,0},e\sscript{0}$, then an entire inspiral can be thought of as traversing a grid of these models, and as such the dynamical friction coefficient, when decoupled from relaxation, can be considered correct. The models in Fig. \ref{L_1.5.pdf} have an asymptotic slope of $\gamma = 1.5$, but similar agreement is found for $\gamma = 1$ and $\gamma = 1.75$.

\subsection{Comparison with GADGET}
\label{gt_discussion.ch}
\subsubsection{Cuspy Models}
The rapid relaxation of low-$N$ models means that even if the dynamical friction force exerted on the cluster is correct at any epoch (as shown in section \ref{nb6_angmom.ch}), different cluster realisations will diverge in agreement due to the stochasticity of the relaxation process. As such, one would ideally like to perform NBODY6 runs with higher particle number to reduce this effect, but the computational cost is too high at the time of writing. Simulation nb1k took 7 days to run on 4 GeForce GTX 780 GPUs and 16 CPU cores. Increasing the particle number by a factor of 10 would take over 2 years to compute. As an alternative we used the softened tree-code GADGET to simulate a larger particle number. We would like to stress that in agreement with \citet{b8}, we cannot accurately describe the internal dynamics of these clusters with GADGET. However tidal stripping still occurs in a natural way, and we can compare the bulk properties (i.e. mass and position) in order to test the validity of our dynamical friction perturbation.

\begin{figure}
 \includegraphics[width=\linewidth]{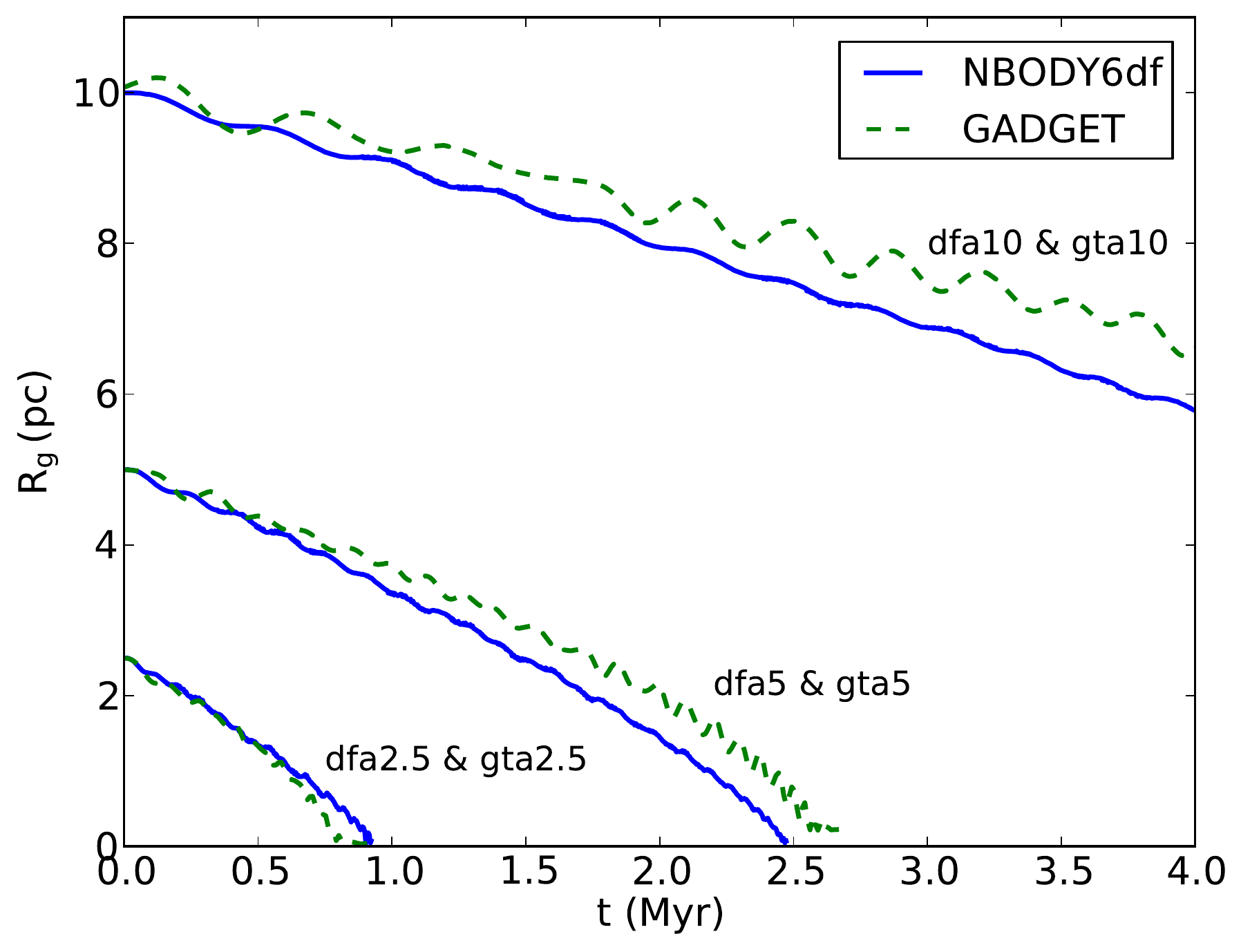}
 \caption{Distance of the cluster with respect to the Galactic Centre as a function of time for NBODY6df (blue line) and GADGET (green dashed line) simulations.} 
 \label{gt_Rg.pdf}
\end{figure}

Softened simulations are expected to exhibit slower dynamical friction than collisional simulations. Whilst softening helps with numerical stability and computational speed, it suppresses the close interactions required for dynamical friction. It is true that the GADGET simulations have a greater mass resolution for the background than nb1k, yet the eccentricity of the orbit grows faster due to numerical inaccuracies in the integration. GADGET's integrator is only accurate to 2nd order, and so accumulates the errors discussed in Section \ref{nb6_discussion.ch} faster than NBODY6's integrator, which is accurate to 4th order. The tree's force calculation is also not as accurate as direct summation, with smoothing effects causing the potential to deviate from spherical symmetry. This deviation is evident in Fig. \ref{gt_Rg.pdf}, which shows the inspiral of cluster models at 10, 5 and 2.5 parsecs in NBODY6df and GADGET. The correspondence is much better between simulations dfa2.5 and gta2.5 because the number density of the background at $2.5\unit{pc}$ is approximately 30 times greater than at $10\unit{pc}$, so the simulation effectively has a higher resolution background, and appears more spherical when calculated with the tree. With this taken into account, the agreement between the NBODY6df and GADGET is rather good. To see if the discrepancy is due to low-$N$, we ran gta5 with twice the mass resolution of the background. The results were indistinguishable and we can be convinced we converged on the solution.

\begin{figure}
 \includegraphics[width=\linewidth]{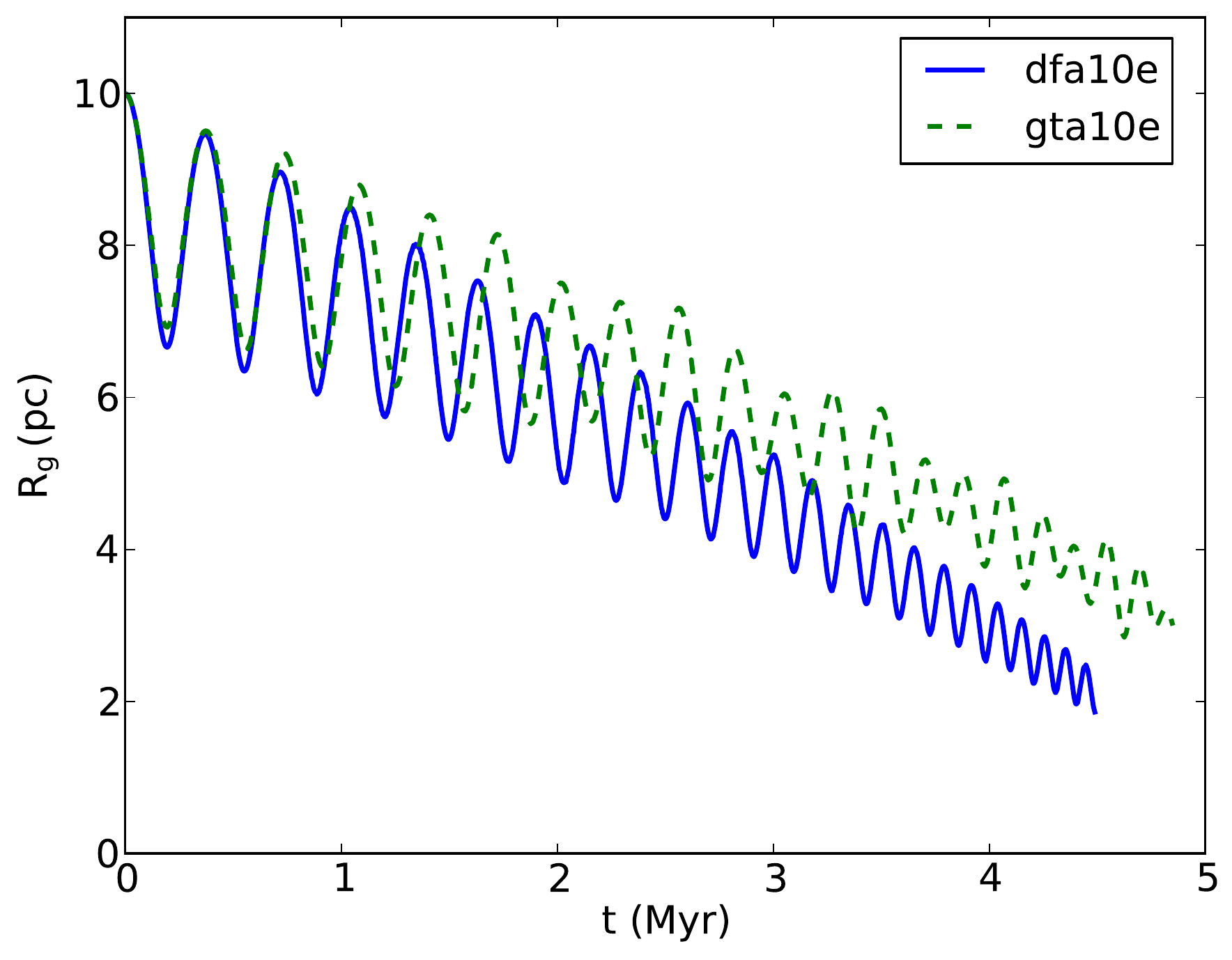}
 \caption{Distance of the cluster with respect to the Galactic Centre as a function of time for dfa10e (blue line) and gta10e (green dashed line).} 
 \label{10k_e03.fig}
\end{figure}

Fig. \ref{10k_e03.fig} shows simulations with the same initial conditions as dfa10 and gta10 in Fig. \ref{gt_Rg.pdf}, but with an initial eccentricity of 0.3. The same general trend is seen as compared with the circular case, in which GADGET inspirals slower, but the first few orbits give very good agreement.

\subsubsection{Shallow Models}
\label{gt_shallow.ch}

In section \ref{theory.ch} we discussed how Chandrasekhar's dynamical friction formula has proven to fail in shallow cusps (i.e. when $\gamma$ approaches 0). The reason for this failure is due to the assumption that dynamical friction is a local process, thus when $\rho(r)$ is constant, no special treatment occurs. \citet{b24,b39} and \citet{b35} show that when orbiting within the scale radius of a shallow profile, satellites experience an enhancement of the dynamical friction force, followed by an abrupt stalling near the core.

The enhanced force is thought to occur due to super-resonance with orbits throughout the core \citep[for the link between Chandrasekhar's formula and the resonant nature of dynamical friction see also][]{b27,b28}. This causes the satellite to interact with each background star more than once, leading to greater dynamical friction as compared with Chandrasekhar's derivation, in which the satellite interacts with each background star only once. It should again be noted that this is why Chandrasekhar's formula works so well for cuspy profiles, where the density is a strong function of $R\sscript{g}$. Whilst migrating to the centre of a cuspy profile the satellite constantly experiences a whole new set of resonances. These new resonances act independently of previous interactions, and as such interacting with any star only once is a reasonable approximation.

\citet{b24} showed that Chandrasekhar's formula with a fixed $\log(\Lambda)$ fails at reproducing inspiral in $N$-body models of a harmonic core. However it may be possible to reproduce these effects with a varying Coulomb logarithm.

\begin{figure}
 \includegraphics[width=\linewidth]{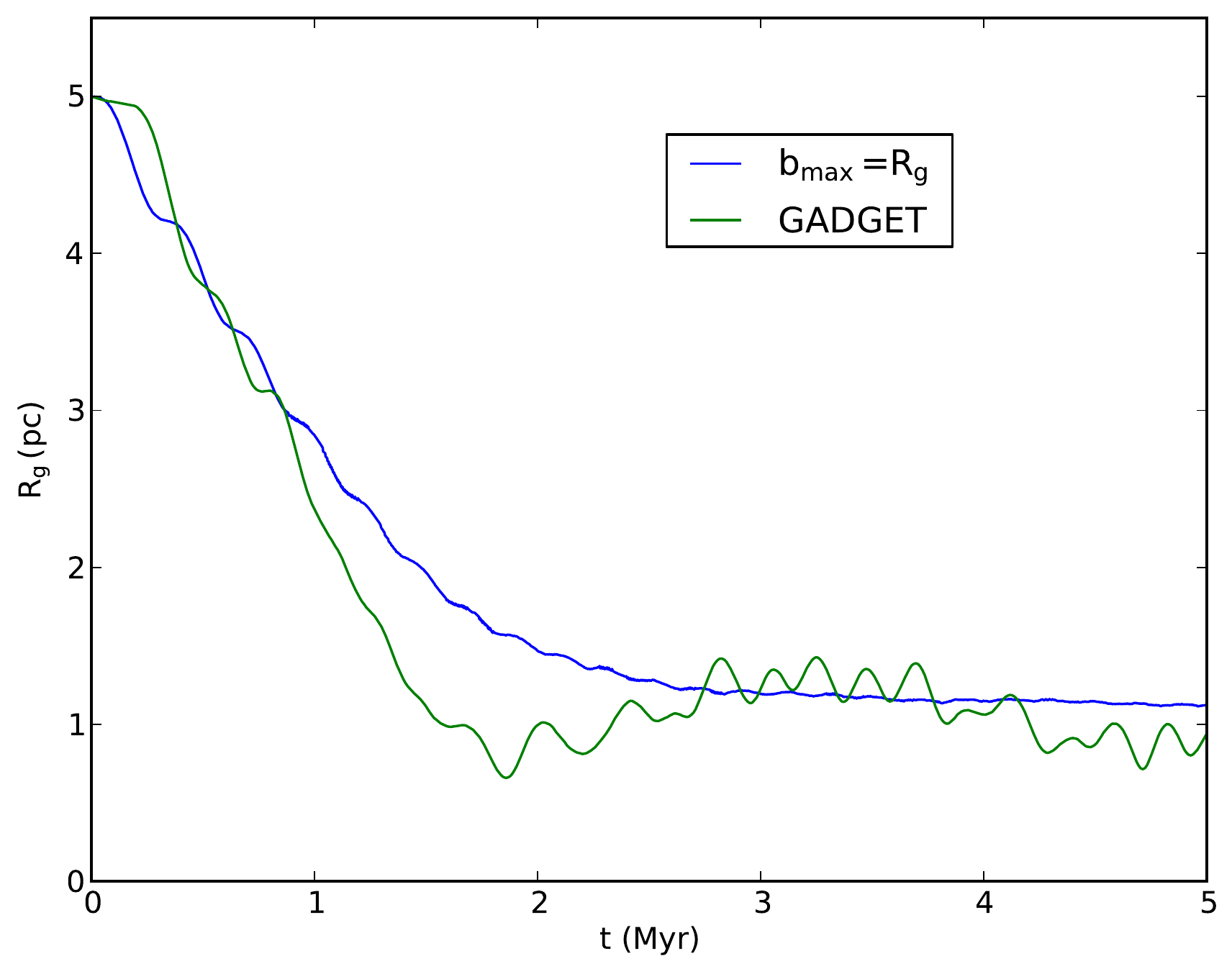}
 \caption{Distance of the cluster with respect to the Galactic Centre as a function of time for dfg0.0 (blue line) and gtg0.0 (green line).} 
 \label{gt_0.fig}
\end{figure}

In Fig. \ref{gt_0.fig} we compare the $\gamma=0$ simulation dfg0.0 with a self consistent GADGET simulation gtg0.0. We choose to only compare GADGET simulations for the shallow models because we cannot compare the full inspiral in NBODY6 due to the fast relaxation effects at low $N$, as seen in previous sections. Studying the full inspiral for the shallow profiles is important as they show interesting deviations from the standard Chandrasekhar's formula.

Fig. \ref{gt_0.fig} shows that this prescription cannot fully reproduce the full inspiral of the cluster during the ``super-Chandraskehar" phase \citep{b24,b35}. The nature of this enhanced force is discussed in section \ref{R_2006.ch}. Interestingly the stalling radius is very well reproduced, thus the physics of the stalling is well captured in this model.

\begin{figure}
 \includegraphics[width=\linewidth]{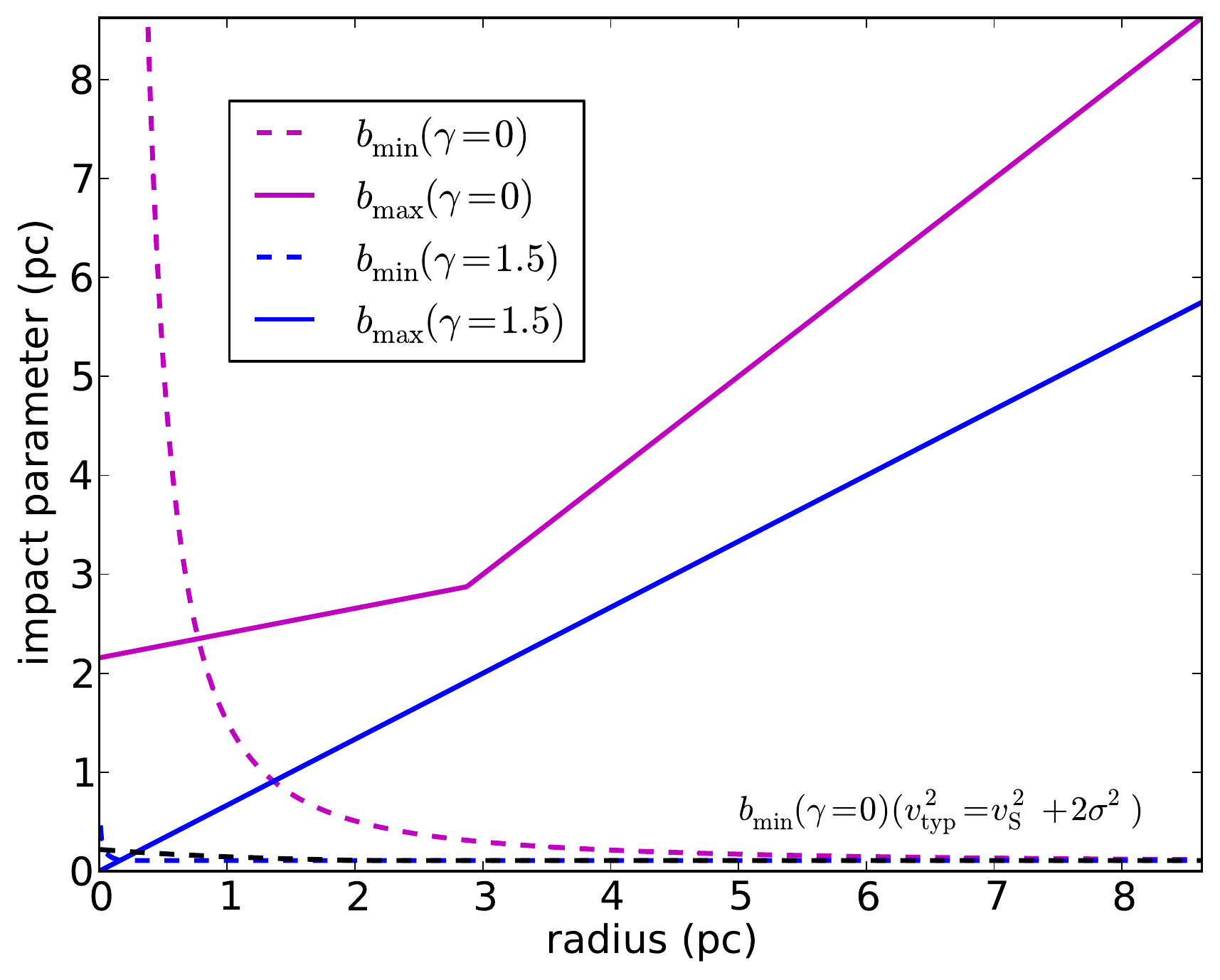}
 \caption{Maximum and minimum impact parameters as a function of radius for $\gamma = 0,1.5$, up to the scale radius. It should be noted that the minimum impact parameter is a function of the mass of the satellite and background, here the values used are the same as throughout this paper. The dashed black line shows the minimum impact parameter if the prescription from \citet{b21} is used for $\gamma=0$.} 
 \label{bmin.fig}
\end{figure}

The cluster stalls rapidly due to a combination of two correlated effects. Firstly, the low circular velocity in the core means that $b\sscript{min}$ is dominated by $GM/v^2\sscript{typ}$ as opposed to the cluster size and becomes very large and comparable to $b\sscript{max}$. The physical interpretation in Chandrasekhar's formalism is that any background particles passing by the cluster are deflected by more than 90 degrees, and thus stochastic changes in orbit due to two-body relaxation dominate over the many-body dynamical friction effect, suppressing any further inspiral. These stochastic changes in orbit are evident in the full $N$-body model. The black dashed line in Fig. \ref{bmin.fig} shows the minimum impact parameter for the $\gamma = 0$ case if one instead defines $v\sscript{typ}^2 = 2\sigma^2 + v^2\sscript{S}$ as in \citet{b22}. It can clearly be seen that the stalling effect is not correctly captured if this prescription is used, as the impact parameters are not equal until $R\sscript{g} \sim 0$. In cuspy profiles most of the mass is contained within a small radius, and stalling occurs very close to the centre of the system.

\begin{figure}
 \includegraphics[width=\linewidth]{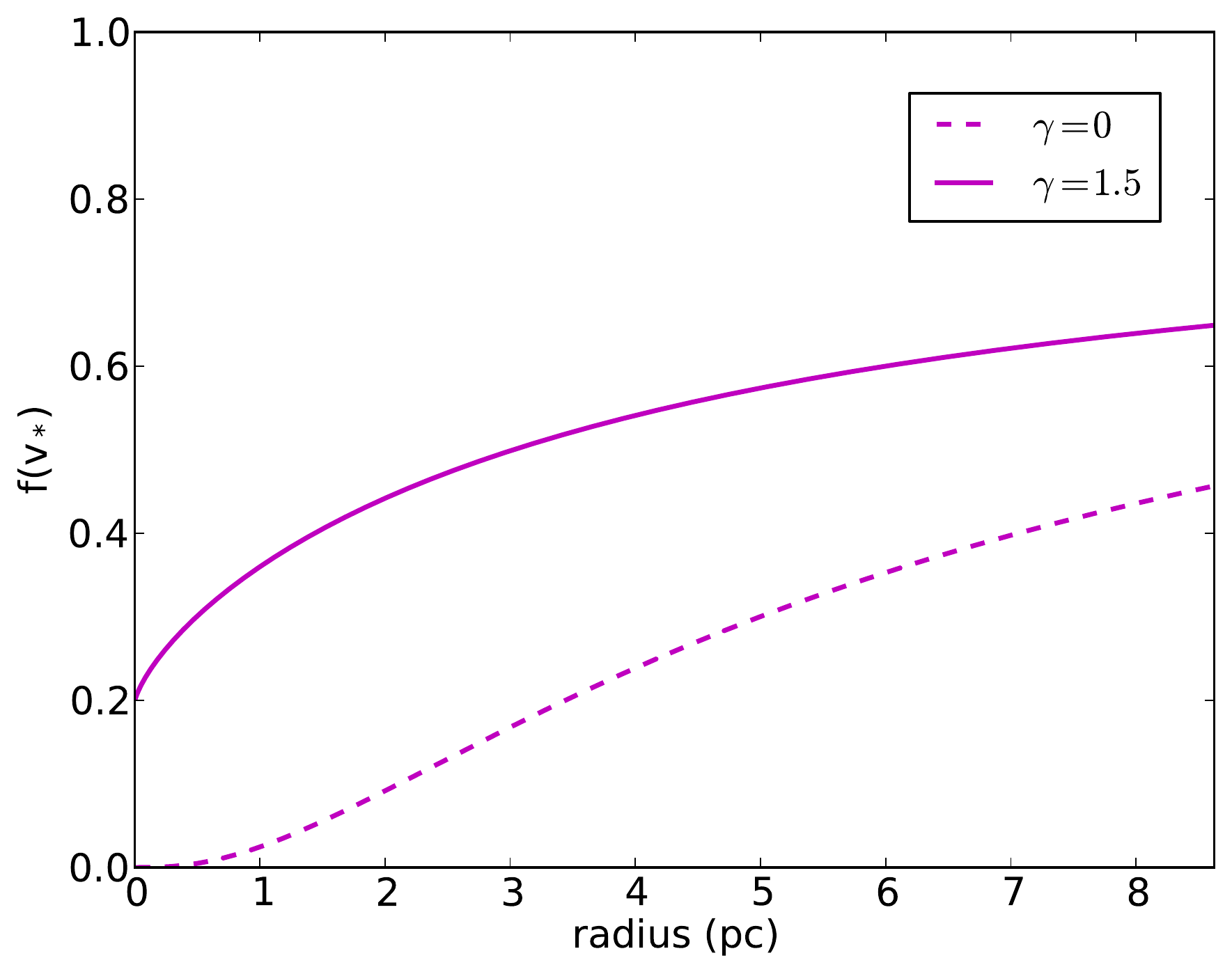}
 \caption{Fraction of stars moving slower than the circular velocity as a function of radius for $\gamma = 0,1.5$. The  distribution of stellar velocities is assumed to be Maxwellian with velocity dispersion $\sigma$, as in equation \ref{dynfric.eq}.} 
 \label{vdist.fig}
\end{figure}

Secondly, as the velocity dispersion doesn't decrease as rapidly as the circular velocity, and is non-zero at $R\sscript{g} = 0$, $X = v\sscript{S}/\sqrt{2}\sigma$ shrinks as $R\sscript{G}$ goes to zero. As such the fraction of stars moving slower than the satellite (those that contribute to dynamical friction) decreases rapidly as the cluster approaches the core. Fig. \ref{vdist.fig} shows the fraction of stars moving slower than the circular velocity as a function of radius for $\gamma =$ 0 and 1.5. In the $\gamma = 0$ case very few stars move slower than the circular velocity in the core, and thus very few stars contribute to dynamical friction. In the cuspy case this fraction of stars does not go to zero, and the stalling effect is dominated by $\log(\Lambda)$ approaching zero, which occurs at very small radii. If the orbit is initially eccentric, faster velocities are reached at pericentre, decreasing $b\sscript{min}$ and causing the satellite to stall closer to the centre of the host, as can be seen in fig. 3 of \citet{b24}.

\begin{figure}
 \includegraphics[width=\linewidth]{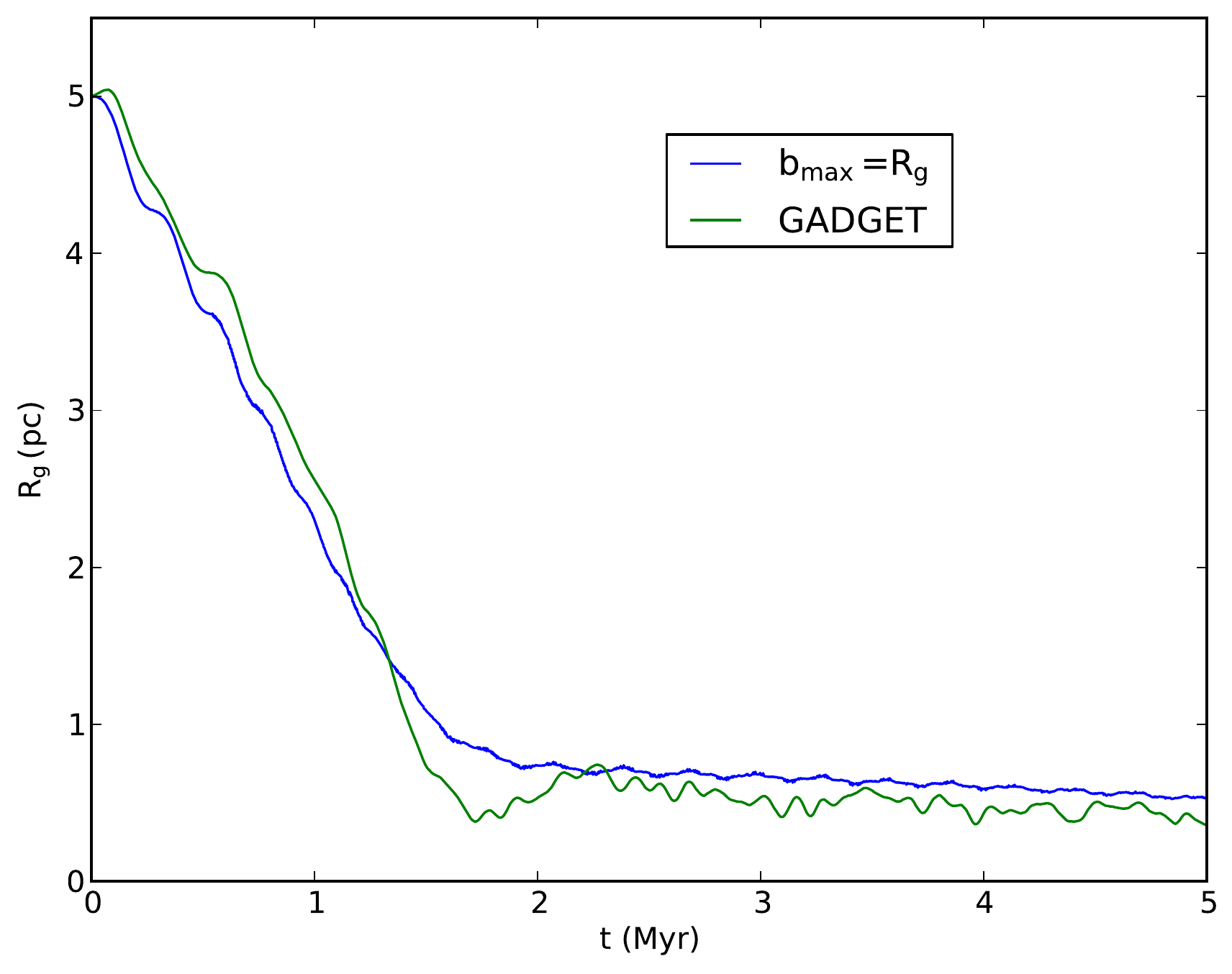}
 \caption{Distance of the cluster with respect to the Galactic Centre as a function of time for dfg0.5 (blue line) and gtg0.5 (green line).} 
 \label{gt_05.fig}
\end{figure}

Fig. \ref{gt_05.fig} shows a similar setup for the $\gamma = 0.5$ profile. which shows excellent agreement with the $N$-body simulation. A cusp of $\gamma = 0.5$ is enough to keep $f(v\sscript{*})$ from going to zero, and the stalling is again dominated by $\log(\Lambda)\rightarrow0$.

It is thus evident that cluster stalling at large radii is characteristic of profiles that contain a constant density core, such that the cluster stalls when:

\begin{equation}
\log(\Lambda)f(v\sscript{*}) \rightarrow 0,
\end{equation}
and dynamical friction ceases when either $M\sscript{S}\rightarrow M\sscript{g}(R\sscript{g})$ or $v\sscript{c} \lesssim \sigma$ (when $v\sscript{c} \sim 0.5\sigma$, $f(v\sscript{*}) \sim 0.03$). As both terms decrease with decreasing satellite velocity, in profiles with a true isothermal core, satellites can stall at $M\sscript{S}/M\sscript{g}(R\sscript{g}) \gg 1$ (see section \ref{R_2006.ch}).

\subsection{N-dependence study with NBODY6df}
\label{nb6_Ndependence.ch}
\begin{figure}
 \includegraphics[width=\linewidth]{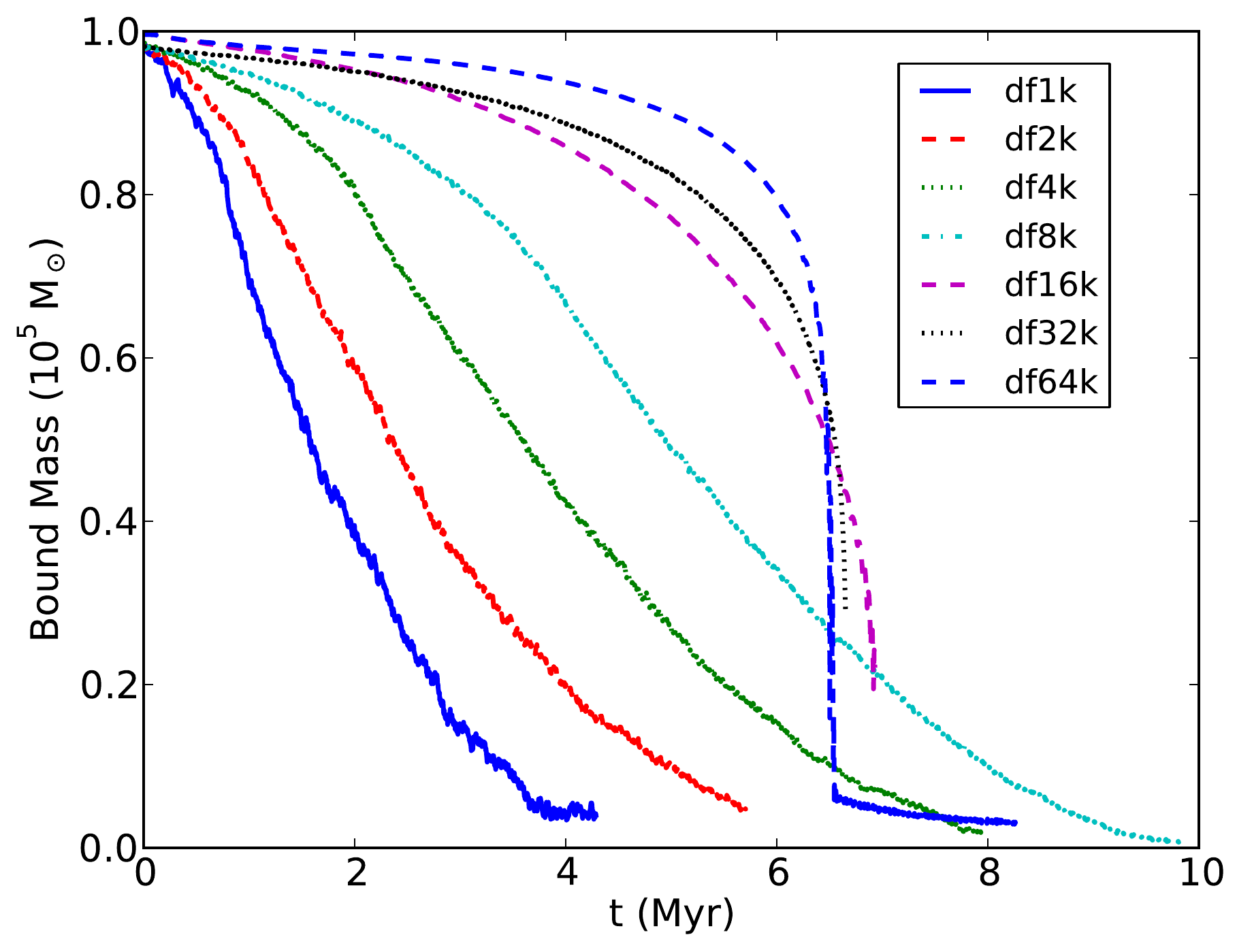}
 \caption{Mass enclosed in the Roche volume as a function of time for simulations df1k-df32k} 
 \label{N_mb.fig}
\end{figure}

We ran a series of simulations to see if the expected $N$-dependence of cluster relaxation, and its effects on inspiral, are well reproduced by NBODY6df. Simulations df1k-df64k have the same initial cluster mass split evenly amongst their cluster particles, and otherwise have the same initial conditions. Low-$N$ systems should lose their mass faster than an equivalent realistic cluster due to shorter relaxation timescales. This behaviour is illustrated in Fig. \ref{N_mb.fig}, which shows the bound mass as a function of time for simulations with different $N$.

With 32k particles the cluster initially has a relaxation timescale of $\sim$ 2Myr, and the mass loss is mostly dominated by the tides during inspiral (Fig. \ref{N_mb.fig}). In simulations df32k and df64k, most of the mass is lost when the cluster reaches the centre of the potential, where the remaining mass is deposited in a disk, and is formally bound to the cusp as opposed to itself.

If the cluster was modelled in a realistic fashion with a mean mass of $0.58 \unit{M_\odot}$ \citep{b18} the relaxation timescale would be longer than the inspiral time ($T\sscript{relax} \sim$ 9Myr), and the mass loss would be dominated by the shrinking tidal radius. The low-$N$ models show accelerated mass loss due to increased dynamical ejections as expected.

\begin{figure}
 \includegraphics[width=\linewidth]{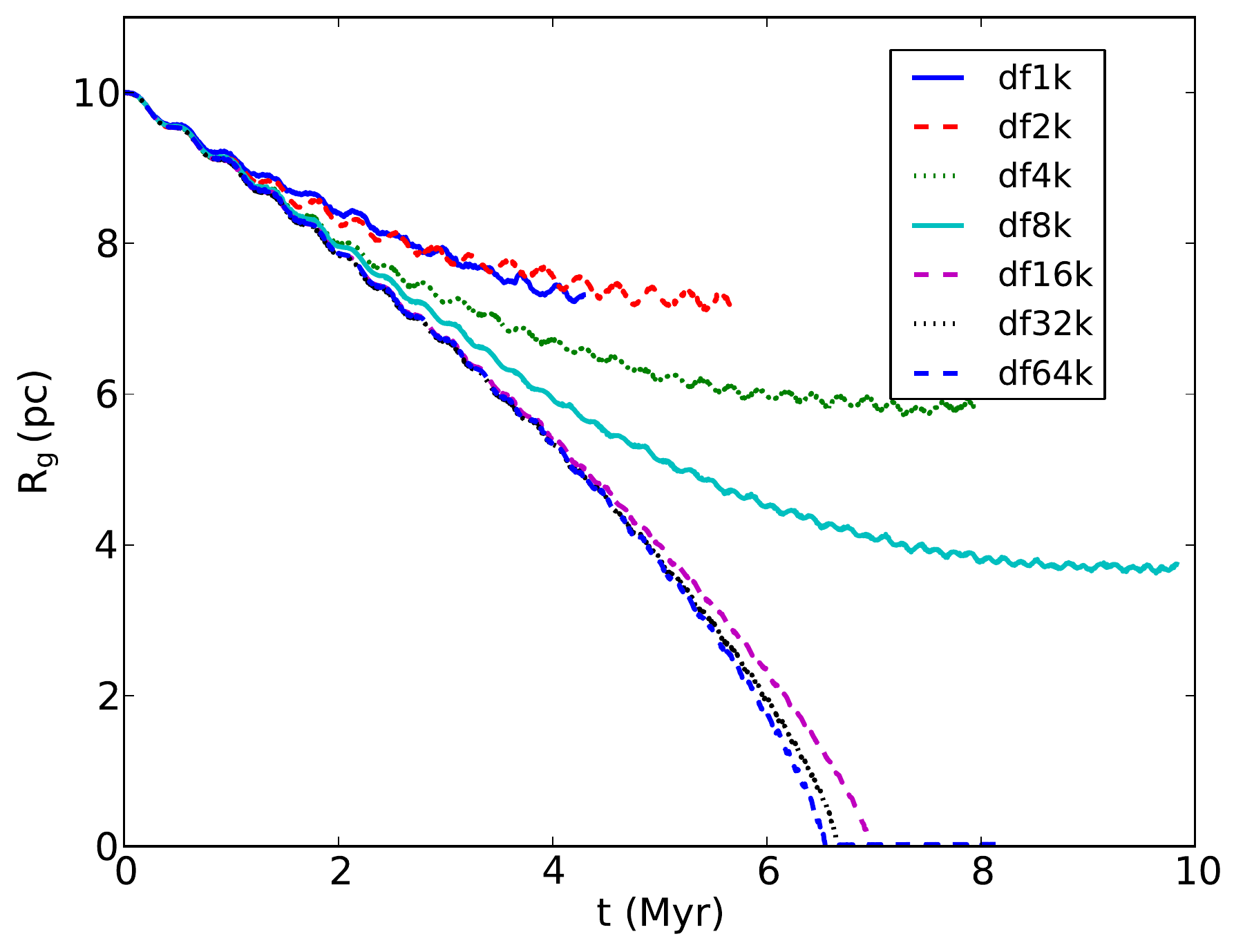}
 \caption{Distance of the cluster with respect to the Galactic Centre as a function of time for simulations df1k-df32k.} 
 \label{N_Rg.fig}
\end{figure}

Fig. \ref{N_Rg.fig} shows how this mass loss drastically alters the evolution of the orbit. If the cluster loses significant mass, its inspiral will stall due to a continually decreasing dynamical friction coefficient.

\FloatBarrier
\section{Discussion}
\label{discussion.ch}

In this section we consider the inspiral of a point mass object, and as such take our dynamical friction formalism and implement it as an external force in an independent 2nd-order integrator, which integrates the motion of a point particle in a Dehnen potential.

\subsection{Understanding stalling in cored profiles}
\label{R_2006.ch}
\begin{figure}
 \includegraphics[width=\linewidth]{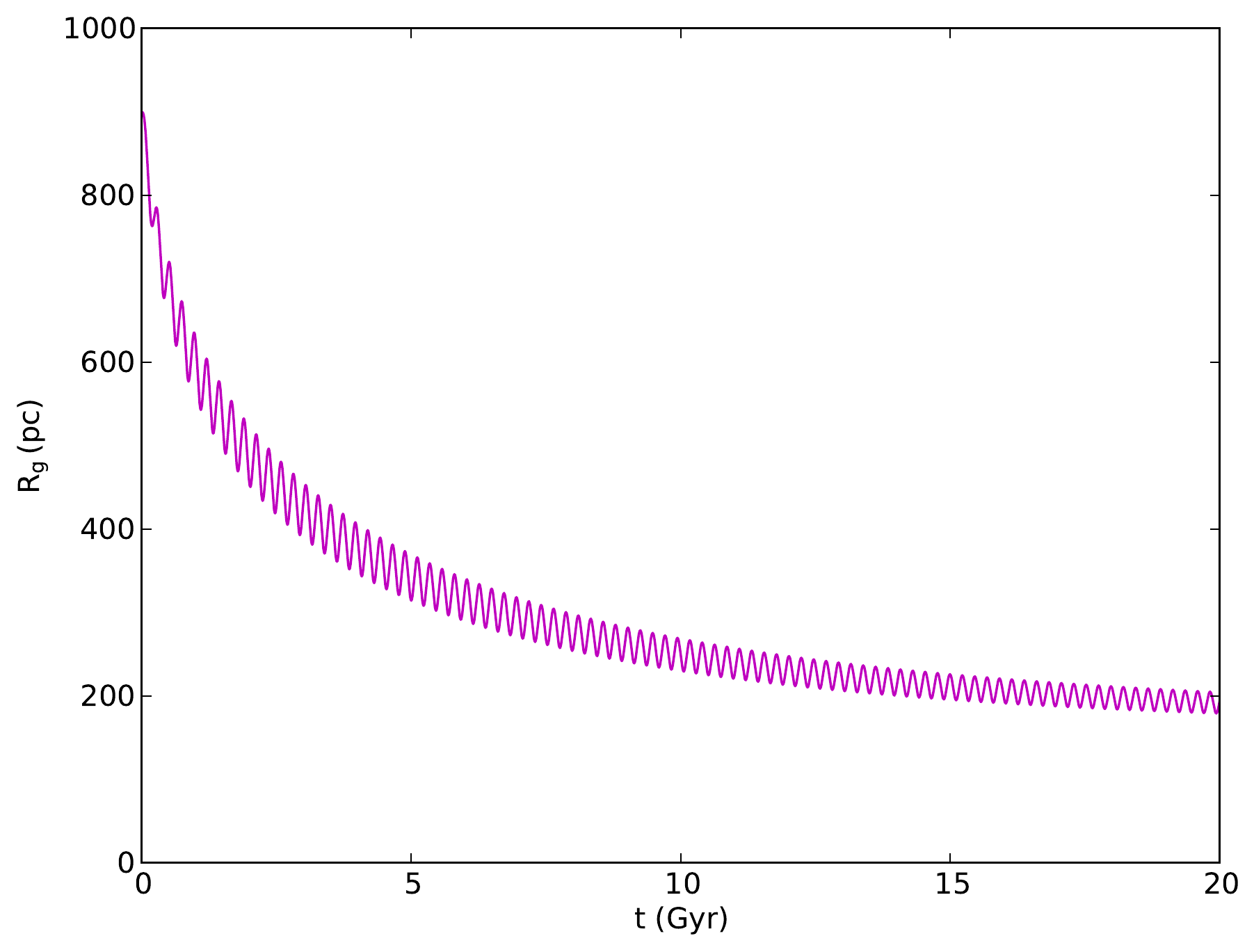}
 \caption{Semi-analytical integration of $2\times10^5\unit{M\sscript{\odot}}$ point-mass cluster in a uniform background density distribution with parameters similar to \citet{b24}.} 
 \label{rstall.fig}
\end{figure}
\citet{b24} studied the stalling of models that contain intrinsic cores using the alpha-beta-gamma profile \citep{b37}:

\begin{equation}
\rho(r) = \frac{\rho\sscript{0}}{(r/a)^{\gamma}(1+(r/a)^{\alpha})^{(\beta-\gamma)/\alpha}},
\label{abg.eqn}
\end{equation}
with $\rho\sscript{0}=9.93\times10^7 \unit{M\sscript{\odot}kpc^{-3}}$, $a=0.91\unit{kpc}$, $\alpha = 1.5$, $\beta=3.0$ and $\gamma=0.0$. The authors could not reproduce the stalling effect with a semi-analytical form of Chandrasekhar's formula, as their model used a constant Coulomb logarithm. In their model the cluster still would have stalled if they had integrated for long enough, due to $f(v\sscript{*}) \rightarrow 0$. However this would have happened at much smaller radii, as they use a constant Coulomb logarithm.

If we consider a toy model where we approximate the inner region of the core as a constant sharply truncated at $a$, the Jeans equation for the velocity dispersion becomes \citep{b2}:
\begin{subequations}
\begin{align}
	v\sscript{r}^2(r) &= \frac{1}{\nu(r)} \int_r^{\infty} \nu(r)\frac{d\Phi}{dr'}\,dr',\\
	\intertext{where $\nu(r)$ is the number density and $\Phi$ is the potential. We can then write:}	
	v\sscript{r}^2(r) &\sim  \int_r^{a} \frac{d\Phi}{dr'}\,dr' \sim \left[\phi\right]^a_r,\\
	v\sscript{r}^2(r) &\sim \left[\frac{GM\sscript{g}(R\sscript{g})}{r}\right]^a_r \sim \left[G\rho\sscript{0}\frac{4}{3}\pi r^2\right]^a_r,\\
	v\sscript{r}^2(r) &\sim G\rho\sscript{0}\frac{4}{3}\pi (a^2-r^2).
	\intertext{Note that as:}
	v\sscript{c}^2 &= \frac{GM\sscript{g}(R\sscript{g})}{r} \sim G\rho\sscript{0}\frac{4}{3}\pi r^2,\\
	\frac{v\sscript{c}}{v\sscript{r}} &= \frac{r}{\sqrt{(a^2-r^2)}},
	\intertext{and:}
	X &= \frac{r}{\sqrt{2(a^2-r^2)}}.
\end{align}
\end{subequations}
Therefore, $v\sscript{c} < v\sscript{r}$ for $r<a/\sqrt{2}$, leading to a quickly vanishing $f(v\sscript{*})$ when approaching the core, causing the cluster to stall much further out. This coupled with the shrinking Coulomb logarithm causes stalling behaviour unique to the core model, in which $M\sscript{g}(R\sscript{g})\gg M\sscript{S}$. In Fig. \ref{rstall.fig} we modelled a $2\times10^5\unit{M\sscript{\odot}}$ point mass cluster in a constant density background with $\rho\sscript{0}$ and $a$ the same as in \citet{b24}. The cluster stalls at $M\sscript{S}/M\sscript{g}(R\sscript{g}) \sim ~0.06$ in agreement with the $N$-body simulations of \citet{b24}. At $\sim200\unit{pc}$, $f(v\sscript{*}) \sim 3\times10^{-3}$, i.e. only 0.3\% of stars move slower than the circular velocity, and dynamical friction practically ceases. Their cluster stalls a little further in as their model follows the distribution of equation \ref{abg.eqn}, and thus $\sigma$ is slightly lower at $\sim200\unit{pc}$.

Our model relies on the key ansatz that dynamical friction ceases when $b\sscript{min} \geq b\sscript{max}$. This can be understood as a lack of particles available to be scattered by less than 90 degrees by the satellite, but the true structure of such a state is not immediately clear. \citet{b24} argued that the system finds itself in a state in which the background particles orbit in the combined potential of the satellite and the harmonic background, in which the time averaged drag force on the satellite can be shown to be zero. The authors showed that this state gives an excellent match to the phase space structure of background stars in the $N$-body simulations. Thus we can understand this configuration being exactly what happens when $b\sscript{min} \geq b\sscript{max}$, and the system remains in a state where no net momentum is transferred from the background to the infalling body.

The "Super-Chandrasekhar" inspiral phase prior to the rapid stalling is currently not captured by our model, however it could possibly arise from the contribution to dynamical friction of the stars moving faster than the satellite. \citet{b38} showed that in shallow cusps around black holes, where most stars are moving faster than the circular velocity, the contribution from the stars moving faster than the satellite is significant. This mechanism is likely to be similar for the case at hand, however it could well be the case that this contribution is not enough to explain the enhanced frictional force, and may result from super-resonance inside the core \citep{b27,b28,b24,b35}. The contribution of the faster stars and their effect on the stalling radius is not trivial, and shall be left to future work.

\subsection{Comparison with \citet{b17}}
\label{AS_CR_2014.ch}

\citet{b17} (hereafter AC14) studied dynamical friction in cuspy galaxies and presented a new treatment for massive objects near the centre of their host systems. The authors derive a semi-analytical formula for the inspiral time of massive point particles orbiting Dehnen models, calibrated by $N$-body models in the GPU-parallel direct $N$-body code HiGPUs \citep{b19}. In their semi-analytic fitting process, they use an exponential interpolation between Chandrasekhar's formula with a constant $b\sscript{max}$ and varying $b\sscript{min}$, and their detailed evaluation of the frictional force near the centre of host systems. The authors do not fix $b\sscript{min}$ and instead let it be a fitting parameter, along with $r\sscript{cr}$, which they define as the critical radius at which they switch to the new regime.

In our model both $b\sscript{max}$ and $b\sscript{min}$ vary along the orbit as a function of the local background and satellite properties. \citet{b17} state that the local approximation overestimates the effects of dynamical friction in the innermost cuspy regions of galaxies, however in our approach the maximum impact parameter tends to zero at small radii for cuspy distributions, reducing the range at which the local approximation acts over. For distributions with $\gamma > 1$ this local scale length is smaller than the distance to the centre of the background. By using $b\sscript{max} = R\sscript{g}/\gamma$ we ensure that the local approximation is valid, but probably slightly underestimate the frictional force at the very centre of the systems, as AC14 suggest \citep[see also][]{b21}.

\citet{b17} model the satellites as Plummer-softened point particles. Simulations with a satellite consisting of a cluster of particles take longer to reach the centre of their host, due to mass loss and the larger $b\sscript{min}$ of extended objects. For this reason we cannot directly test NBODY6df against their timescale formula. Instead we use our semi-analytical integrator. We also perform GADGET simulations with the same initial conditions to compare our results. The list of simulations is presented in Table \ref{point_mass.tbl}.

\begin{table}
 \begin{minipage}{85mm}
  \begin{tabular}{@{}llrrr@{}}
   Simulation & Code&$R\sscript{a}$ & e & $\gamma$\\
   & &($\unit{pc}$) & \\
   \hline
   gtpt1.5 & GADGET & 5.0 & 0.0 & 1.5\\
   gtpt1.5e & GADGET & 5.0 & 0.3 & 1.5\\
   gtpt1.0 & GADGET & 5.0 & 0.0 & 1.0\\
   gtpt1.0e & GADGET & 5.0 & 0.3 & 1.0\\
   sapt1.5 & Semi-Analytic Integrator & 5.0 & 0.0 & 1.5\\
   sapt1.5e & Semi-Analytic Integrator & 5.0 & 0.3 & 1.5\\
   sapt1.0 & Semi-Analytic Integrator & 5.0 & 0.0 & 1.0\\
   sapt1.0e & Semi-Analytic Integrator & 5.0 & 0.3 & 1.0\\
   
   \hline

\end{tabular}
\caption{Simulations in which the cluster is modelled by a point mass particle. Simulations are performed in a self consistent way in GADGET, the mass resolution of the GADGET simulations is $30\unit{M\sscript{\odot}}$. The properties of the cluster and background are the same as in section \ref{IC.ch}.}
\label{point_mass.tbl}
\end{minipage}

\end{table}

We compared the inspiral time of these simulations with results from AC14 and found significant discrepancy with their dynamical friction timescale, which was calibrated mostly on radial orbits (equation 20 in AC14). However a good agreement is found with an improved formula calibrated on a wider range of models, given in \citet{b25} (arXiv:1501.04567).

\begin{figure}
 \includegraphics[width=\linewidth]{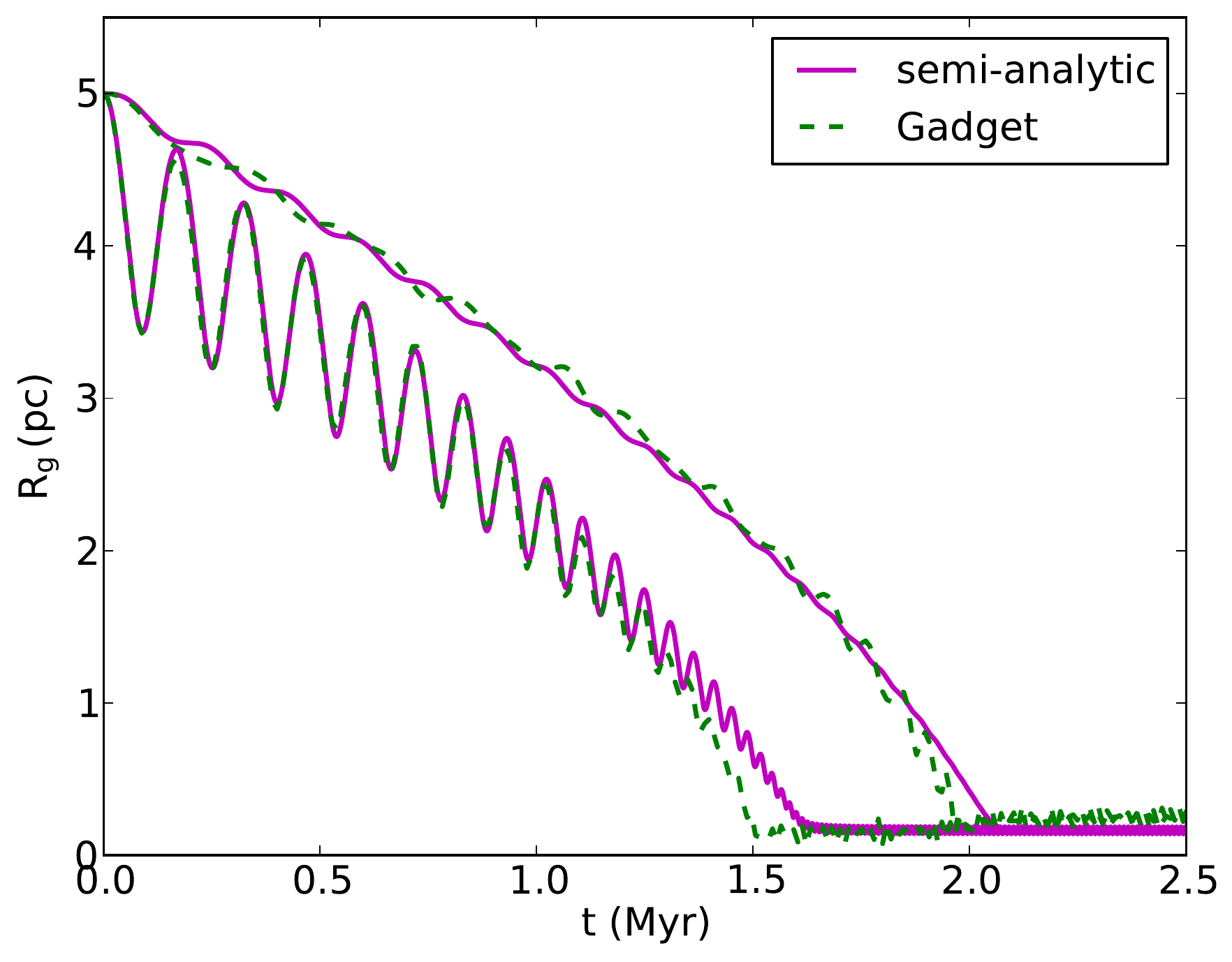}
 \caption{Distance of the cluster with respect to the Galactic Centre as a function of time for simulations gtpt1.5, gtpt1.5e, sapt1.5 and sapt1.5e.} 
 \label{point_15.fig}
\end{figure}

\begin{figure}
 \includegraphics[width=\linewidth]{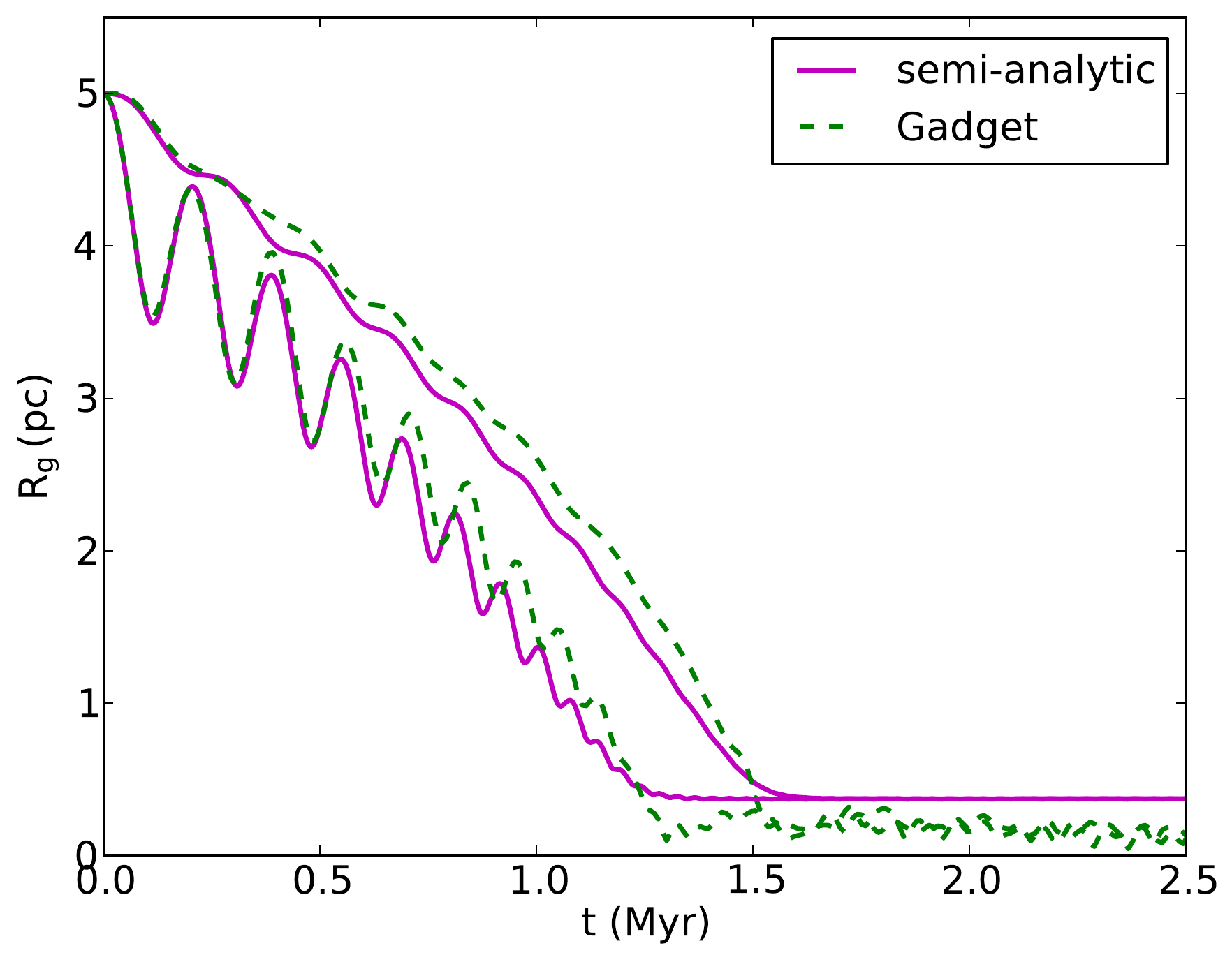}
 \caption{Distance of the cluster with respect to the Galactic Centre as a function of time for simulations gtpt1.0, gtpt1.0e, sapt1.0 and sapt1.0e.} 
 \label{point_1.fig}
\end{figure}

Both our semi-analytic approach and GADGET simulations show good agreement with the revised timescale formula, (see Fig. \ref{point_1.fig} and \ref{point_15.fig} for the $\gamma = 1.0$ and $\gamma = 1.5$ cases, respectively). The radial trajectory of the inspiral in GADGET is very well reproduced by our semi-analytic formula and we can validate our approach for the inner cuspy regions. In the $\gamma = 1.0$ the semi-analytic approach diverges slightly from the GADGET simulation, as the live background distribution is slightly shallowed by the inspiraling body, however the match is still reasonably good, with the inspiral time being well captured. A mechanism for feeding the energy lost by the satellite back into the analytic background distribution would be able to correct for this effect for massive satellites in shallow background distributions, however this is beyond the scope of this work.

It should be noted that gtpt1.5e agrees much better with our semi-analytic model than gta10e does with NBODY6df. This is because the effect of close encounters (i.e. $b\sim b\sscript{min}$) is completely captured by the high mass ratio in the point mass case. Within a particular cluster, the effect of close encounters with background stars at the edge of the cluster are underestimated. The NBODY6 simulations treat these encounters properly, and thus excellent agreement is found.

Chandrasekhar's local approximation is inaccurate near the centre of cuspy host systems, and AC14's approach is more representative of the true force in the very central region of the background distribution. We recommend the use of AC14's numerically calculated Coulomb logarithm when a very accurate representation of inspiral is required in the very inner region of cuspy profiles. However, our varying Coulomb logarithm can reasonably approximate the force experienced throughout the cuspy region, only slightly underestimating the inspiral time, without the need for two free parameters.

\section{conclusions}
\label{conclusion.ch}
We present  a modification to the GPU-enabled direct summation code NBODY6, which we name NBODY6df, to include the effects of dynamical friction on the inspiral of a star cluster in a smooth background particle distribution. In this approach, the dynamical friction force on each cluster particle is computed analytically and added to the $N$-body forces exerted by the other cluster particles. In this way, only the cluster needs to be modelled in a $N$-body fashion, while the effect of the background stars is included in an approximated but reliable way. This significantly reduces computational time with respect to a full $N$-body modelling of the cluster and the background system.

It should be emphasised that the dynamical friction treatment implemented in NBODY6df is physically motivated rather than calibrated on $N$-body simulations, and thus has predictive power, owing to the physically motivated maximum and minimum impact parameters. The predictive power of NBODY6df allows for quick modelling of a large parameter space of initial conditions without prior calibration. The mass term in Chandrasekhar's formula for extended objects is found to be well represented by the mass enclosed in the Roche volume, as opposed to just the formally bound stars. This is due to the presence of potential escapers enhancing the gravitational wake whilst they are still close to the cluster. NBODY6df can be used to simulate young cluster inspiral in the Galactic Centre, or the inspiral of globular clusters or dwarf galaxies in the halo of a larger host. It should be noted that dynamical friction in a disk or other highly non-spherical systems cannot yet be reliably modelled with NBODY6df. This is due to the maximum impact parameter being smaller perpendicular to the disk than parallel to it. Accurately modelling inspiral in a disk would require an angular dependence in the summation of possible impact parameters, and is beyond the scope of this work. A comprehensive study of young dense clusters formed close to the Galactic Centre will be presented in an upcoming paper. 

NBODY6df can also accurately model the inspiral of satellites in shallow profiles, due to a new approach in which the minimum parameter is defined to be inversely proportional to the square of the satellite's velocity, and independent of the velocity dispersion.

For a direct summation code, computational time scales with $N^2$ for an integration of one $N$-body time unit. A full $N$-body simulation of a $10^5 \unit{M_{\odot}}$ cluster with mean mass of $0.58\unit{M_{\odot}}$ and a Kroupa mass function  would require $\sim 9 \times 10^7$ background particles for a 10:1 ratio of $M\sscript{bg}: M\sscript{c}$. On the other hand, a simulation with NBODY6df would only require the $1.73\times 10^5$ cluster particles to be modelled as $N$-body particles, reducing the computational time by several orders of magnitude.

While the current implementation adopts a Dehnen model for the background system, any static model can be implemented in order to follow the evolution of star clusters in which dynamical friction of the orbit is important. The code will be released publicly on github\footnote{https://github.com}.

\section*{Acknowledgements}

We thank the anonymous referee for suggestions which helped improve the scientific content of this paper. Simulations were run on the GPU cluster hosted by the Astrophysics group at the University of Surrey. We thank Mark Gieles for his consistent help with technical issues regarding NBODY6, and Elisa Bortolas for the initial conditions of Dehnen models. We also thank Douglas Heggie for his help regarding technical issues that arose when attempting to run large-$N$ simulations in NBODY6. JIR would like to acknowledge support from STFC consolidated grant ST/M000990/1 and the MERAC foundation.

\bibliography{biblio}

\appendix
\renewcommand{\theequation}{A\arabic{equation}}
\section*{appendix A: Velocity dispersion}
Here we derive the velocity dispersion as a function of radius for Dehnen models. We also include the optional potential of a central black whole, which although not used in this paper, is available in the initial release of NBODY6df.

The velocity dispersion for an isotropic spherical system is \citep{b2}:

\begin{equation}
	v\sscript{r}^2(r) = \frac{1}{\nu(r)} \int_r^{\infty} \nu(r)\frac{d\Phi}{dr'}\,dr'
	\label{Jeans.eqn}
\end{equation}
Where $\nu(r)$ is the number density and $\Phi$ is the potential. The number density of a Dehnen model is given by:

\begin{equation}
	\nu(r) = \frac{\rho(r)}{M\sscript{g}}=\frac{(3-\gamma)}{4\pi} \frac{a}{r^\gamma (r+a)^{4-\gamma}}
	\label{numdens.eqn}
\end{equation}
And the potential by \citep{b13}:
\begin{equation}
	\Phi(r) = -\frac{GM\sscript{g}}{a}\frac{1}{2-\gamma}\left[ 1 - \frac{r}{r+a}^{2-\gamma}  \right] - \frac{G\mu M\sscript{g}}{r}
	\label{pot.eqn}
\end{equation}
Where the second term is the additional potential due to an optional central black hole, and $\mu$ is the ratio of the black hole and Dehnen model masses. It follows that: 
\begin{equation}
	\frac{d\Phi}{dr} = \frac{GM\sscript{g}\left(\frac{r}{r+a}\right)^{1-\gamma}}{(a+r)^2} + \frac{G\mu M\sscript{g}}{r^2}
	\label{dpot.eqn}
\end{equation}
By putting equations \ref{numdens.eqn} and \ref{dpot.eqn} into equation \ref{Jeans.eqn} and making the substitution $x=r/a$, equation \ref{Jeans.eqn} becomes:
\begin{equation}
	v\sscript{r}^2(x) = \frac{GM\sscript{g}}{a}x^{\gamma}(x+1)^{4-\gamma}\left[ f(x) + \mu h(x)\right]
	\label{sigma.eqn}
\end{equation}
Where:
\begin{equation}
	f(x) = \int_x^{\infty} \frac{\left(\frac{x}{x+1}\right)^{1-\gamma}}{x^{\gamma}(x+1)^{6-\gamma}}  \,dx
	\label{fx.eqn}
\end{equation}
\begin{equation}
	h(x) = \int_x^{\infty} x^{-\gamma-2}(x+1)^{\gamma-4} \,dx
	\label{fx.eqn}
\end{equation}
Which must be evaluated for the desired value of $\gamma$. $f(x)$ and $h(x)$ are analytic for integer values of $4\gamma$. If $\mu$ is zero the $h(x)$ term is skipped in NBODY6df.

\label{lastpage}

\end{document}